\documentclass[aps,twocolumn,pra,superscriptaddress,amsmath,showpacs,tightenlines]{revtex4-1}
\usepackage{bbm}
\usepackage{amssymb}
\usepackage{amsmath}
\usepackage{graphicx}
\usepackage{subfigure}
\usepackage{natbib}
\usepackage{epsfig}
\usepackage{amsfonts}
\usepackage{mathrsfs}
\usepackage{CJK}
\usepackage{xcolor}
\usepackage[toc,page,title,titletoc,header]{appendix}

\begin{document}
\title{Two-photon scattering in a waveguide by a giant atom}
\author{Yang Xue}
\affiliation{School of integrated circuits, Tsinghua University, Beijing 100084, China}
\author{Yu-xi Liu}
\email{yuxiliu@mail.tsinghua.edu.cn}
\affiliation{School of integrated circuits, Tsinghua University, Beijing 100084, China}
\affiliation{Frontier Science Center for Quantum Information, Beijing, China}

\begin{abstract}
We study two-photon scattering  by a two-level giant atom in a waveguide. We first study the case that the giant atom is coupled to the waveguide via two coupling points, and obtain Bethe ansatz eigenstates and eigenvalues in the Hilbert space of two-excitation. Then we derive bound states by subtracting the states corresponding to Bethe ansatz solutions from the entire two-excitation Hilbert space, and construct the two-photon scattering matrix (S-matrix) by using Bethe ansatz eigenstates and bound states. We further study the properties of output states, which include both the scattering and bound states, for arbitrarily incident two-photon states by using a concrete example. We find that  the oscillation period of the scattering states and  decay rates of the bound states strongly depend on the distance between two coupling points. Moreover, we find that the two-photon correlation in the bound states can be enhanced by changing such distance when the total energy of two incident photons equals to two times of single photon resonance energy. We also generalize our study to the case that the giant atom is coupled to the waveguide via $N$ coupling points. We obtain all the eigenstates and eigenvalues of the scattering matrix and construct the S-matrix. Comparing with the case of the two coupling points, we find the photon correlation can be further enhanced by increasing the number of the coupling points for the same incident states when the distance of any two nearest neighbor coupling points is half of the wavelength.
\end{abstract}
\maketitle
	
\section{Introduction}
The light-matter interaction plays a crucial role in quantum optics and quantum information processing. It includes absorption, emission, transmission, reflection and scattering of light. In past years, there is a growing interest in the research on the interaction between atoms and quantized electromagnetic fields in waveguide~\cite{Roy2017RMP,Gu2017PR,Sheremet2023RMP},  which is an indispensable element for quantum information transmission in quantum networks~\cite{Kimble2008N}. Single-photon scattering in the waveguide, by single natural or artificial atoms~\cite{Shen2005OL,Shen2005PRL}, many atoms~\cite{Tsoi2008PRA,Liao2015PRA},  or single-photon atomic mirrors~\cite{Chang2012NJP,Mirhosseini2019N}, has been studied.  Moreover,  many theoretical methods have been developed to study the scattering of two or more photons, e.g., real-space Bethe ansatz method~\cite{Shen2007PRA,Shen2007PRL,RoyPRB2010,RoyPRL2011,Roy2013PRA}, Lehmann-Symanzik-Zimmermann reduction approach~\cite{Shi2009PRB,Shi2011PRA}, the path integral~\cite{Shi2015PRA}, wave-packet evolution~\cite{Liao2010PRA,Liao2013PRA}, Lippmann-Schwinger (LS) formalism~\cite{Zheng2013PRL,Fang2014EPJ,Roy2011PRA}, and input-output theory~\cite{Fan2010PRA,Rephaeli2011PRA}.

Recently, the interaction between giant atoms and quantized electromagnetic fields in the waveguide has emerged as a new research subject. Unlike natural atom or usual artificial atom, which is idealized as a point particle and coupled to the waveguide via single spatial point,  the giant atom has big size compared to wavelengths of electromagnetic fields in the waveguide, thus it cannot be considered as a point particle and is coupled to the waveguide via several spatial points. Experimentally, superconducting quantum circuits~\cite{Kockum2019Springer,Blais2021RMP} acting as giant atoms have been coupled to surface acoustic waves~\cite{Gustafsson2014S}, in which the wavelengths of the surface acoustic waves are comparable to the size of the giant atom. Moreover, the coupling of superconducting artificial atoms to microwave waveguides has also been demonstrated through multiple coupling points~\cite{Kannan2020N,Vadiraj2021PRA}. The non-local multiple-point coupling of the giant atom to the waveguide can result in interference of electromagnetic fields or surface acoustic waves. This multiple point coupling induced interference may lead to various novel phenomena compared to natural atoms, such as non-exponential decay~\cite{Guo2017PRA,Andersson2019NP}, frequency-dependent decay rates and Lamb shifts~\cite{Kannan2020N}, decoherence-free interactions between braided giant atoms~\cite{Kockum2018PRL,Carollo2020PRR,Du2023PRA}, enhanced generation of entangled states~\cite{Santos2023PRL}, oscillating bound states~\cite{Guo2020PRR}.
	
Single-photon scattering by a single giant atom~\cite{Cai2021PRA} or more giant atoms~\cite{FengPRA2021,Yin2022PRA,Peng2023PRA} in the waveguide has been studied. Two-photon correlations have also been explored when two incident photons are scattered by one or more giant atoms in the waveguide with LS formalism under Markov approximation~\cite{Gu2023PRA,Gu2024PRA}. It was found that the  intrinsic interference can enhance the control of two-photon bound states. Meanwhile, non-Markovian single-photon and two-photon scattering by giant atoms has also been studied~\cite{Gu2024PRA2}. Moreover, the photon statistics for two-photon scattering by a giant Kerr cavity without the Markovian approximation has also been studied~\cite{Chang2024arxiv}.
	
Despite significant advances have been made in two-photon scattering by giant atoms in the waveguide~\cite{Gu2023PRA,Gu2024PRA}, the eigenstates and corresponding eigenvalues of the scattering matrix (S-matrix)  have not been given.  Here, we first derive eigenstates and corresponding eigenvalues of the  two-photon S-matrix by using the real-space Bethe ansatz method within the Markovian approximation~\cite{Cheng2023LP}, in which the giant atom is coupled to  the waveguide via two spatial points. These eigenstates can be used to construct the scattering matrix as shown in Ref.~\cite{Shen2007PRA} for natural atom,  and are applied to study the photon-photon correlation of the scattering light in waveguide by natural atoms~\cite{Mahmoodian2018PRL,Iversen2021PRL}. We then obtain the S-matrix of the two-photon scattering for giant atom with two coupling points and further apply the S-matrix to study the properties of the two-photon scattering for a given incident two-photon state. Moreover, we extend our study to the case that the giant atom is coupled to the waveguide via $N$ spatial points, which has not been explored by others to our best knowledge.

The paper is organized as follows: In Sec.~\ref{sec:model},  the theoretical model is introduced. In Sec.~\ref{sec:single},  the single-photon S-matrix is derived, the transmission and reflection coefficients of the single-photon are given. In Sec.~\ref{sec:two}, the eigenstates and eigenvalues of S-matrix are given by using Bethe ansatz method under the Markovian approximation, and the two-photon S-matrix is given.  In Sec.~\ref{sec:two mode}, we derive output state for a given incident two-photon state by using S-matrix and analyze the properties of the output state. We also compare the results with those of the two-photon scattering by natural atom. In Sec.~\ref{sec:Nleg}, we study two-photon scattering by a giant atom with $N$ coupling points to the waveguide and corresponding S-matrix is derived. We compare the properties of the output state for two-photon scattering by the giant atom of $N$ coupling points with those of two coupling points for the same incident two-photon state. Finally, we summarize our results and discuss the possible experimental realization in Sec.~\ref{sec:conclusion}.

\section{Model}\label{sec:model}
	
\begin{figure}
\centering
\includegraphics[width=0.9\linewidth]{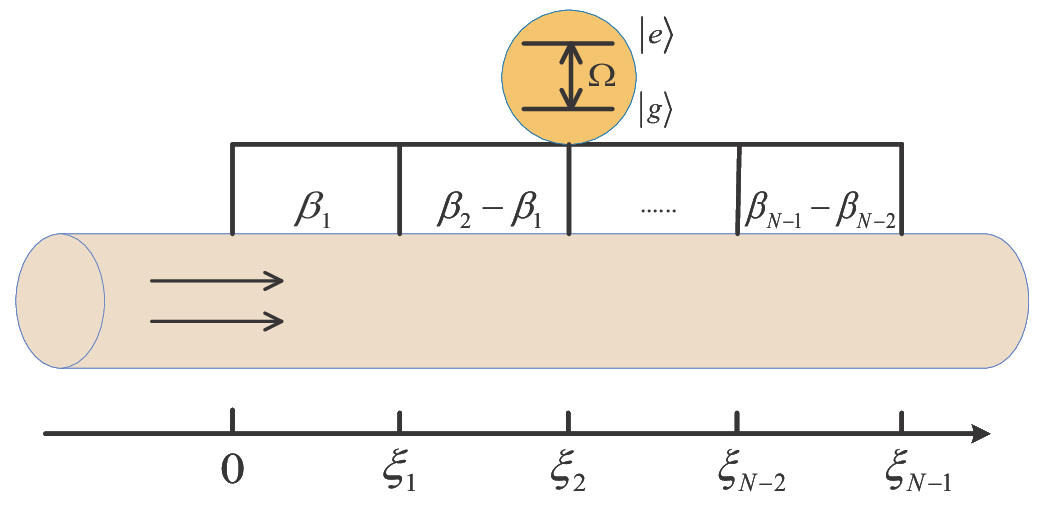}
\caption{A schematic diagram for a giant atom coupled to a waveguide via $N$ coupling points at the positions $\xi_0=0$, $\xi_1$, $\xi_2$, $\ldots$, $\xi_{N-1}$. The equivalent phases between the nearest neighbor coupling points are $\beta_1=k_0 (\xi_1-\xi_0)$, $\beta_2 - \beta_1=k_0 (\xi_2-\xi_1)$, $\ldots$, $\beta_{N-1} - \beta_{N-2}=k_0 (\xi_{N-1}-\xi_{N-2})$. Here, the Markov approximation has been adopted with $k_0=\Omega$ where the velocity of photons $v_g=1$.}
\label{fig:fig1}
\end{figure}

As schematically shown in Fig.~\ref{fig:fig1}, we study a system that a giant atom is coupled to a waveguide via $N$ spatial points.  The Hamiltonian of the system can be written as
\begin{align}\label{eq:sysN}
H &= iv_g\int dx \left[ C_L^\dagger(x) \frac{\partial}{\partial x} C_L(x) - C_R^\dagger(x) \frac{\partial}{\partial x} C_R(x) \right] \nonumber\\
&\quad +\frac{\sqrt{2}V}{N}\sum_{i=0}^{N-1} \int dx \, \left[ \delta(x - \xi_i)\right] \left[ C_R^\dagger(x) \sigma_-  \right. \nonumber\\
&\quad \left. +C_L^\dagger(x) \sigma_- + \text{H.c.} \right] + \Omega |e\rangle \langle e|.
\end{align}
Here, we set $\hbar=1$ and the energy of the ground state of the giant atom as zero. The parameter $v_g$ denotes the propagating velocity of the photons in the waveguide. $C_{L}$ and $C_R$ are the annihilation operators for $L$-mode and $R$-mode photons, which propagate to the left and right, respectively. The coupling strength $\sqrt{2}V/N$ between the giant atom and the waveguide at the coupling points $x = \xi_0,\,\,\xi_1,\,\,...,\,\,\xi_{N-1}$ is the same. Hereafter, we set $\xi_0=0$ and also use angle to denote these positions of the coupling points, e.g., $\beta^{\prime}_{i}=k\xi_i$ with the wave vector $k$.  $\sigma_+=|e\rangle\langle g|$ and $\sigma_-=|e\rangle\langle g|$ represent the raising and lowering operators of the giant atom with the transition frequency $\Omega$ from the ground $|g\rangle$  to the excited $|e\rangle$ state.

For simplicity and without loss of generality, we first study the case that the giant atom is coupled to the waveguide via two coupling points and then extend the discussions to the case of $N$ coupling points in Sec.~\ref{sec:Nleg}. For the case of two coupling points, the Hamiltonian in Eq.~(\ref{eq:sysN}) is reduced to
\begin{align}
H &= iv_g\int dx \left[ C_L^\dagger(x) \frac{\partial}{\partial x} C_L(x) - C_R^\dagger(x) \frac{\partial}{\partial x} C_R(x) \right] \nonumber\\
&\quad +\frac{V}{\sqrt{2}} \int dx \, \left[\delta(x) + \delta(x - \xi_1)\right] \left[ C_R^\dagger(x) \sigma_-  \right. \nonumber\\
&\quad \left. +C_L^\dagger(x) \sigma_- + \text{H.c.} \right] + \Omega |e\rangle \langle e|. \label{eq:sys}
\end{align}
If we define $e$-mode and $o$-mode operators as
\begin{align}
C_e^\dagger(x) &\equiv \frac{1}{\sqrt{2}} \left[ C_R^\dagger(x) + C_L^\dagger(-x) \right], \nonumber\\
C_o^\dagger(x) &\equiv \frac{1}{\sqrt{2}} \left[ C_R^\dagger(x) - C_L^\dagger(-x) \right], \label{eq:dec_eo}
\end{align}
via the $L$-mode and $R$-mode operators,	then  the Hamiltonian in Eq.~(\ref{eq:sys}) can be rewritten as  $H =  H_o+H_e$ with
\begin{equation}\label{eq:eq6}
H_o =-i v_g \int \mathrm{d}x \left[ C_o^\dagger(x) \frac{\partial}{\partial x} C_o(x) \right]
\end{equation}
and
\begin{align}
H_e=H_0+H_{\text{int}}\label{eq:eq3}
\end{align}
with
\begin{align}
H_0 &= \Omega |e\rangle \langle e|  -i  v_g \int \mathrm{d}x \left[ C_e^\dagger(x) \frac{\partial}{\partial x} C_e(x) \right],\label{eq:H_0} \\
H_{\text{int}}&= \frac{V}{2} \int \mathrm{d}x \, \left[ \delta(x) + \delta(x - \xi_1) \right] \left[ C_e^\dagger(x) \sigma_- + \text{H.c.} \right].\label{eq:H_int}
\end{align}
 Equations~(\ref{eq:eq6}) and (\ref{eq:eq3}) clearly show  that the giant atom interacts only with the $e$-mode of the waveguide. Photons of the $o$-mode propagate freely as plane waves. Thus, it is sufficient to only calculate the S-matrix in the $e$-mode space.

\section{Single photon S-matrix and scattering}\label{sec:single}

To study two-photon scattering, we need first to calculate the S-matrix of the single-photon scattering. As shown in Eqs.~(\ref{eq:eq6}) and (\ref{eq:eq3}),  the single-photon scattering is mainly determined by the interaction between the $e$-mode and the giant atom, and the $o$-mode photons freely evolve, thus we need only to derive the single-photon S-matrix $S_e$ for $e$-mode photons by solving the eigenstates and eigenvalues in the space of the single-excitation corresponding to the Hamiltonian in Eq.~(\ref{eq:eq3}). In the single-excitation space,  the eigenstate $\exp(-iEt)\left|k^{(e)}\right\rangle$ corresponding to the eigenvalue $E= v_g k$ of the $e$-mode Hamiltonian $H_e$ in Eq.~(\ref{eq:eq3})  has following form
\begin{equation}\label{eq:E}
\left|k^{(e)}\right\rangle = \int \mathrm{d}x [g(x) C_e^\dagger(x) + e_k \sigma_+] |0, g\rangle,
\end{equation}
which satisfies the eigenvalue equation $H_e |k^{(e)}\rangle = E |k^{(e)}\rangle$. Where the state $|0, g\rangle$ denotes that the waveguide is in the vacuum state $|0\rangle$ and the giant atom is in its ground state $|g\rangle$.  The superscript ``(e)"  denotes the $e$-mode. Hereafter, we call the eigenstate $\left|k^{(e)}\right\rangle$  as interacting state. The parameter $g(x)$ represents the probability amplitude that there is an $e$-mode photon in the waveguide and the giant atom is in the ground state $|g\rangle$, and $e_k$ represents the probability amplitude that the giant atom is in the excited state $|e\rangle=\sigma^\dagger|g\rangle$ and the waveguide is in the vacuum state $|0\rangle$. By solving the eigenvalue equation $H_e |k^{(e)}\rangle = E |k^{(e)}\rangle$, the relation between coefficients  $g(x)$ and $e_k$  is given as
\begin{eqnarray}
&\left(-i v_g \frac{\partial}{\partial x} - E\right) g(x) + \frac{V}{2} \left[\delta(x) + \delta(x - \xi_1)\right] e_k = 0, \label{eq:gx} \\
&\frac{V}{2} [g(0) + g(\xi_1)] + (\Omega - k) e_k = 0. \label{eq:ek}
\end{eqnarray}
Equation~(\ref{eq:gx}) indicates that $g(x)$ is a piecewise continuous function. Without loss of generality, we assume that $g(x)$  has the following form
\begin{equation}
g(x) = e^{ikx} \begin{cases}
1 & \text{for } x < 0, \\
f_k & \text{for } 0 < x < \xi_1, \\
t_k & \text{for } x > \xi_1.
\end{cases} \label{eq:ansatz}
\end{equation}
Here, $ f_k $ represents the probability amplitude of a photon being in the waveguide between the two coupling points, and $ t_k $ represents the transmission coefficient of the single-photon in the $e$-mode space. Substituting Eq.~(\ref{eq:ansatz}) into Eq.~(\ref{eq:gx}) and Eq.~(\ref{eq:ek}), we obtain
\begin{align}
f_k &= \frac{4\Delta_k}{4\Delta_k + iV^2 [1 + \exp(ik \xi_1)] }, \\
t_k &= \frac{4\Delta_k - iV^2 [1 + \exp(-ik \xi_1)] }{4 \Delta_k + iV^2 [1 + \exp(ik\xi_1)]}, \label{eq:tk1} \\
e_k &= \frac{2V [1 + \exp(ik\xi_1)] }{4\Delta_k + iV^2 [1 + \exp(ik \xi_1)] },
\end{align}
with the detuning $\Delta_k = k - \Omega$ between the photon with the wave vector $k$ and the giant atom, where we have used the relations $g(0) = [g(0^-) + g(0^+)] / 2$ and $g(\xi_1) = [g(\xi_1^-) + g(\xi_1^+)] / 2$.
	
The eigenstate $|k^{(e)}\rangle$ in Eq.~(\ref{eq:E}) and the remote past (or future) state $|k_i^{(e)}\rangle$ (or $|k_f^{(e)}\rangle$) satisfy the Lippmann-Schwinger equations~\cite{Sakurai,Shen2007PRA}.
\begin{align}
\left|k^{(e)}\right\rangle&=\left|k^{(e)}_i\right\rangle+\frac{1}{E-H_0+i\epsilon}H_{\text{int}}\left|k^{(e)}\right\rangle, \label{eq:LS1}\\
\left|k^{(e)}\right\rangle&=\left|k^{(e)}_f\right\rangle+\frac{1}{E-H_0-i\epsilon}H_{\text{int}}\left|k^{(e)}\right\rangle. \label{eq:LS2}
\end{align}
Here, $|k^{(e)}_i\rangle$ describes the photon eigenstate that the interaction between the giant atom and the photons is switched off in the far left and remote past with $x<<0$ and $t\rightarrow-\infty$, thus it is governed by the free Hamiltonian in Eq.~(\ref{eq:H_0}). Similarly,   $|k^{(e)}_f\rangle$ describes the photon eigenstate that the interaction between the giant atom and the photons is switched off in the far right and remote future with  $x>>\xi_{1}$ and $t\rightarrow\infty$.
It has been proved~\cite{Sakurai} that these states $|k^{(e)}_i\rangle$, $|k^{(e)}_f\rangle$ and $|k^{(e)}\rangle$ satisfy the eigenvalue equation with the same energy $E$, i.e., $H_0|k^{(e)}_i\rangle=E|k^{(e)}_i\rangle$, $H_0|k^{(e)}_f\rangle=E|k^{(e)}_f\rangle$ and $H_e|k^{(e)}\rangle=E|k^{(e)}\rangle$.  Thus, all eigenstates $|k^{(e)}_i\rangle$ or  $|k^{(e)}_f\rangle$ corresponding to the Hamiltonian $H_{0}$ in the remote past or future  can be constructed by  the interacting state $|k^{(e)}\rangle$ via Eq.~(\ref{eq:LS1}) or Eq.~(\ref{eq:LS2}). To distinguish the states $|k^{(e)}_i\rangle$  and $|k^{(e)}_f\rangle$ with the state  $|k^{(e)}\rangle$, we call the $|k^{(e)}_i\rangle$ as ``in-state" and $|k^{(e)}_f\rangle$ as ``out-state" as in Ref.~\cite{Shen2007PRA}. They can be obtained as $|k^{(e)}_i\rangle= \int dx \,e^{ikx} C_e^\dagger |0,g\rangle/\sqrt{2\pi}$ and $|k^{(e)}_f\rangle=t_k|k^{(e)}_i\rangle $ by substituting $|k^{(e)}\rangle$ in Eq.~(\ref{eq:E}) into Eqs.~(\ref{eq:LS1}) and (\ref{eq:LS2}).

From the definition of the S-matrix~\cite{Sakurai}, we can derive the S-matrix $S_e=\sum_k|k^{(e)}_f\rangle\langle k^{(e)}_i|$ of the $e$-mode in the single excitation space as
\begin{eqnarray}
S_e &= \sum t_k \left|k_i^{(e)}\right\rangle \left\langle k_i^{(e)}\right|, \nonumber \\
\end{eqnarray}
where the summation is taken over the complete basis $|k_i^{(e)}\rangle$ of the Hilbert space determined by the Hamiltonian $H_0$ given in Eq~(\ref{eq:H_0}). The scattering matrix $S_o$ for the $o$-mode is the identity matrix because the Hamiltonian $H_o$ in Eq.~(\ref{eq:eq6}) describes the free propagation of photons in the $o$-mode space. It can be written as
\begin{equation}
S_o = \sum  \left|k_i^{(o)}\right\rangle\left\langle k_i^{(o)}\right|
\end{equation}
with $|k_i^{(o)}\rangle= \int dx \,e^{ikx} C_o^\dagger |0,g\rangle/\sqrt{2\pi}$. Thus, the total scattering matrix of the single-photon can be written as
\begin{equation}\label{eq:18}
S=S_e+S_o.
\end{equation}

Thus, for any incident single-photon state, we can use the S-matrix in Eq.~(\ref{eq:18}) to derive an output state.  As an example, let us now derive the scattered output state in the $R$-mode and $L$-mode spaces by giving an incident  state $|k^{(R)}_1\rangle= \int dx e^{ik_1x} C_R^\dagger |0,g\rangle/\sqrt{2\pi} $ of the $R$-mode in the real space. We can rewrite the state $|k^{(R)}_1\rangle$ into the superpositions
\begin{align}
\left|k^{(R)}_1\right\rangle &=\frac{1}{\sqrt{2\pi}} \int dx e^{ik_1x} C_R^\dagger |0,g\rangle \nonumber \\
&= \frac{1}{2\sqrt{2\pi}}\int dx e^{ik_1x}\left[ C_e^\dagger |0,g\rangle +  C_o^\dagger |0,g\rangle\right] \nonumber \\
&= \frac{1}{\sqrt{2}}\left( \left|k_1^{(e)}\right\rangle +  \left|k_1^{(o)}\right\rangle\right),
\end{align}
 of states $|k_1^{(e)}\rangle$ and $|k_1^{(o)}\rangle$ in the $e$-mode and $o$-mode spaces. The output state $|k_{\rm out}\rangle$ corresponding to the incident state $|k^{(R)}_1\rangle $ can be obtained by applying the S-matrix in Eq.~(\ref{eq:18}) as
\begin{align}
|k_{\rm out}\rangle &=S \left|k_1^{(R)}\right\rangle = \frac{1}{\sqrt{2}} \left[S_e \left|k_1^{(e)}\right\rangle + S_o \left|k_1^{(o)}\right\rangle\right] \nonumber \\
&= \frac{1}{2} (t_{k_1} + 1) \left|k_1^{(R)}\right\rangle + \frac{1}{2} (t_{k_1} - 1) \left|-k_1^{(L)}\right\rangle \nonumber \\
&= \bar{t}_{k_1} \left|k_1^{(R)}\right\rangle + \bar{r}_{k_1} \left|-k_1^{(L)}\right\rangle,
\end{align}
where the transmission coefficients $t_{k} $ of $e$-mode and $o$-mode spaces can be expressed by the transmission $\overline{t}_{k}$ and reflection $\overline{r}_{k}$ coefficients of the $R$-mode and $L$-mode spaces as
\begin{align}
\overline{t}_{k} &= \frac{t_{k} + 1}{2} = \frac{4\Delta_k - i\Gamma \sin(\beta_1') }{4\Delta_k + i\Gamma [1 + \exp(i\beta_1')] }, \nonumber \\
\overline{r}_{k} &= \frac{t_{k} - 1}{2} = \frac{-i\Gamma (1 + \cos(\beta_1')) }{4\Delta_k + i\Gamma [1 + \exp(i\beta_1')] }, \label{eq:tk}
\end{align}
which coincides with Ref.~\cite{Cai2021PRA}. Here, we define the decay rate $\Gamma = V^2$ and the phase $\beta_1' = k \xi_1 = (1 + \Delta_k / \Omega) \beta_1$ with  the phase difference $\beta_1 = k_0 \xi_1$ between the coupling points $0$ and $\xi_1$, and $k_0=\Omega/v_g$. In the single-excitation case, equation~(\ref{eq:tk}) is an exact solution, which is applicable in both the Markovian approximation and the non-Markovian case.
	
We can compare the differences of the single-photon transmission coefficients in the Markovian and non-Markovian regimes. For the giant atom, we define the time delay between two coupling points as $T = \xi_1 / v_g\equiv  \xi_1$ for $v_g=1$, and the relaxation time is $1 / \Gamma$. In the Markovian regime, the time delay should be shorter than the relaxation time, i.e., $T \ll 1 / \Gamma$. Under this condition, we can replace the detuning-dependent phase factor $\beta_1'$ in Eq.~(\ref{eq:tk}) with a detuning-independent phase factor $\beta_1$. This approximation requires that $\beta_1 \Delta_k / \Omega \ll 1$, which is valid in the Markovian approximation with $\beta_1 / \Omega \ll 1 / \Gamma$, given that the bandwidth $\Delta_k$ of interest is on the order of $\Gamma$~\cite{Cai2021PRA}. Consequently, we obtain the reflection coefficients in the Markovian approximation as
\begin{equation}
R=|\overline{r}_{k}|^2=\frac{\Gamma^2[1+\cos(k_0 \xi_1)]^2}{(4\Delta_k-\Gamma \sin(k_0 \xi_1))^2+\Gamma^2[1+\cos(k_0 \xi_1)]^2}\label{eq:R}
\end{equation}
	
\begin{figure}
\centering
\includegraphics[width=0.48\linewidth]{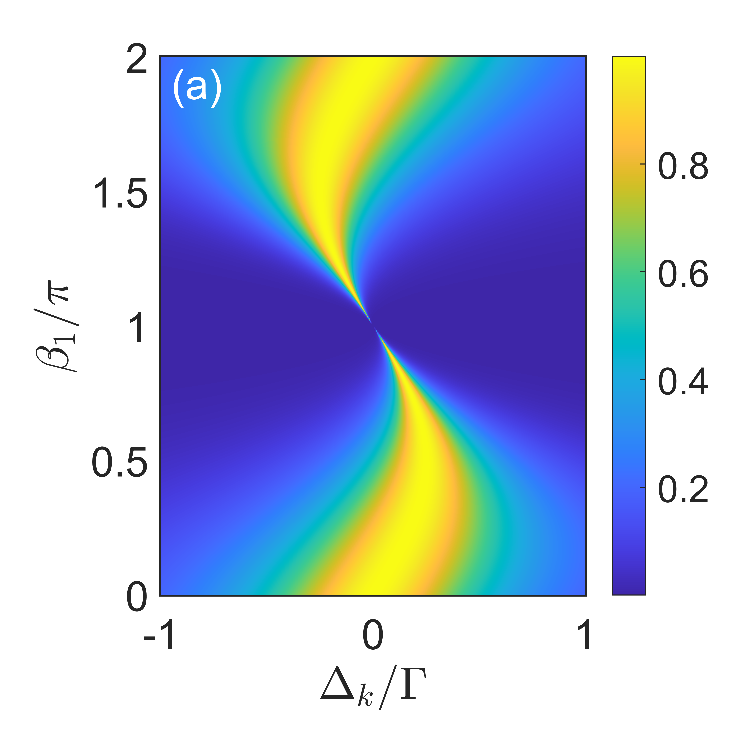}
\includegraphics[width=0.48\linewidth]{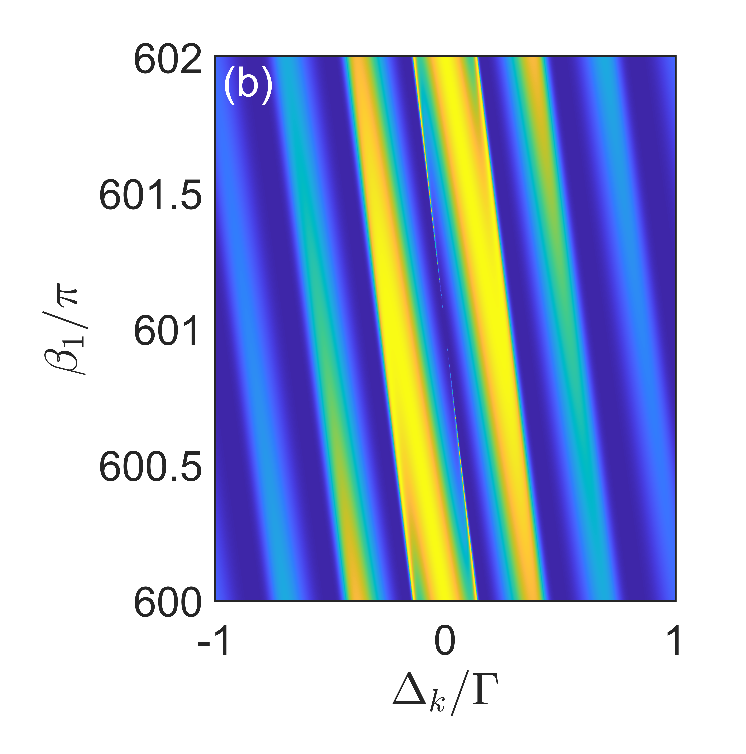}
\caption{The reflection coefficient $\overline{r}_{k}$ as a function of the detuning $\Delta_k$ and the phase $\beta_1 = k_0 \xi_1$ is shown for (a) in  the Markovian approximation and (b) in the non-Markovian case. Here, we set $V^2 = \Gamma$ and $\Omega = 100 \Gamma$.}
\label{fig:fig2}
\end{figure}
	
In Fig.~\ref{fig:fig2}(a), we plot the reflection coefficient as a function of detuning $\Delta_k$ and phase $\beta_1 = k_0 \xi_1$. Here, we use the Markovian approximation by replacing $k$ with $k_0$, which satisfies the relation $k_0=\Omega$ for $v_g=1$.  In contrast to the natural atom, where the reflection coefficient of hundred percent occurs only at resonance, i.e., $\Delta_k = 0$,  we find that the reflection coefficient of hundred percent for the giant atom with two coupling points depends on both the detuning $\Delta_k$ and the phase $\beta_1$, as shown in Fig.~\ref{fig:fig2}(a). From Eq.~(\ref{eq:R}), it is clear that the the peak position of the maximum reflection coefficient is determined by  both the Lamb shift $\Delta_L$ and the detuning $\Delta_k$, while the peak width represents the effective decay rate $\Gamma_g$.
\begin{align}
\Delta_L&=\frac{\Gamma \sin(\beta_1)} {4},\nonumber\\
\Gamma_g&=\frac{1+\cos(\beta_1)}{2}\Gamma.\label{eq:gammag}
\end{align}
That is, the resonant condition of the maximum reflection for the giant atom is changed to $\Delta_k=\Delta_L$ from $\Delta_k=0$ for the natural atom.

As shown in Fig.~\ref{fig:fig2}(b), the relation $T \ll 1/\Gamma$ is no longer valid for a larger value of $\beta_1$ in the non-Markovian case. With the variations of $\Delta_k/\Gamma$ and $\beta_1$, we find that the reflection coefficient $|\overline{r}_{k}|^2$ displays an oscillatory behavior with the variations of $\Delta_k$, rather than the Lorentzian line shape. For example, at $\beta_1 = 600\pi$, three peaks appear near $\Delta_k = 0$, as discussed in Ref.~\cite{Cai2021PRA}. This is because the accumulated phase $\beta_1 \Delta_k / \Omega$ in the free space cannot be neglected  for a larger value of $\beta_1$ in the non-Markovian case. This phase leads to a detuning-dependent value of $\beta_1'$ and results in oscillatory behavior of  the reflection coefficient $|\overline{r}_{k}|^2$.

\section{Two-photon S-matrix} \label{sec:two}

In this section, we construct the S-matrix for two-photon scattering in the two-excitation space.  As discussed in Sec~\ref{sec:model},  the $o$-mode is decoupled from the giant atom. Thus, we need only to calculate eigenstates in the two-excitation space corresponding to the Hamiltonian in Eq.~(\ref{eq:eq3}), which describes the interaction between the $e$-mode and the giant atom. Using these two-excitation eigenstates, we can obtain the in-states and out-states as studied in Sec~\ref{sec:single} by using  Lippmann-Schwinger equations~\cite{Sakurai,Shen2007PRA}, and thus construct the S-matrix for two-photon scattering.
\begin{figure}
\centering
\includegraphics[width=0.8\linewidth]{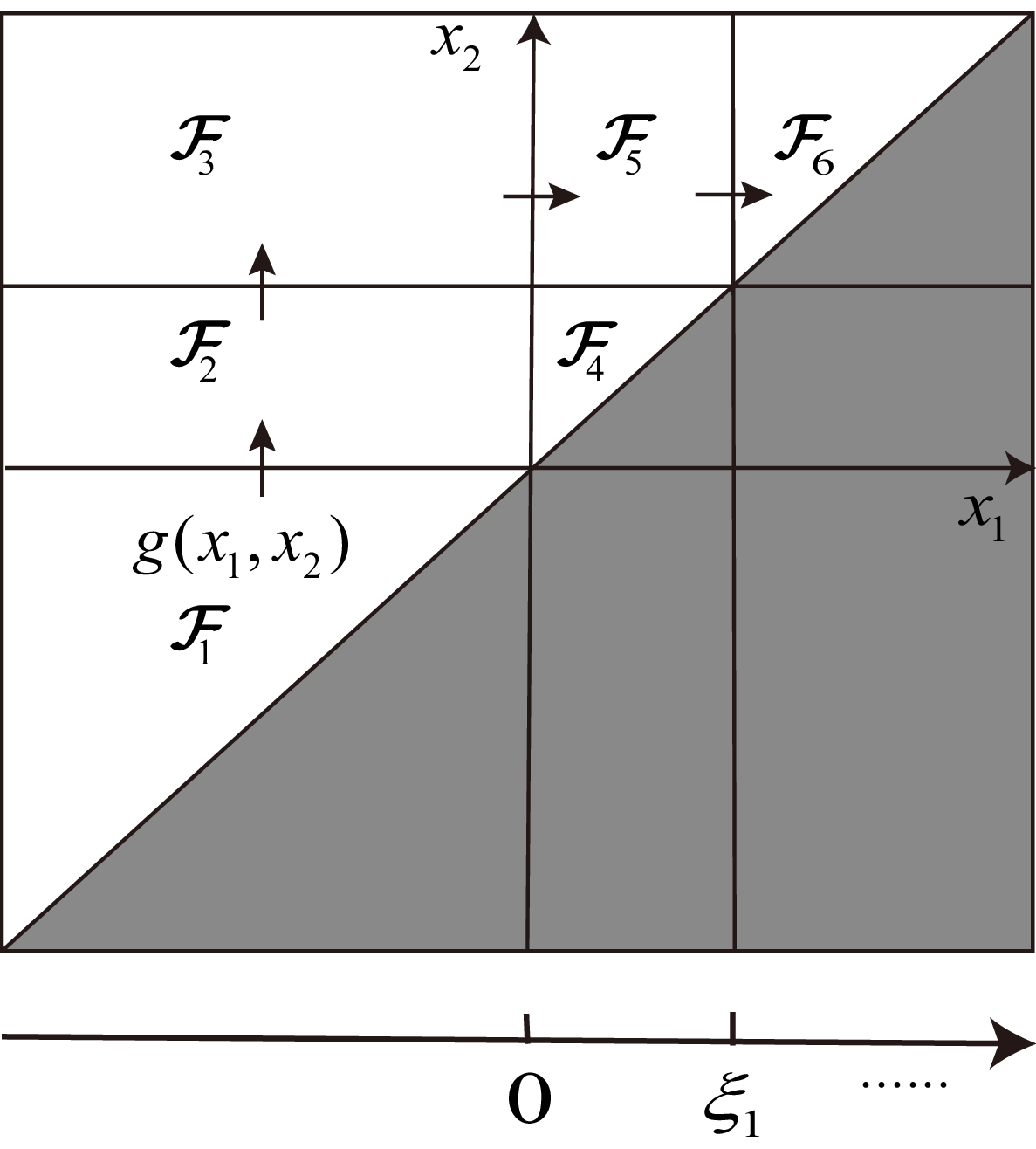}
\caption{The coordinate axes $x_1$ and $x_2$ divide the upper half-plane (unshaded areas) into six regions, labeled from $\mathcal{F}_1$ to $\mathcal{F}_6$, by two coupling points. With three coupling points, the regions extend similarly. The unshaded areas correspond to the case where $x_2 > x_1$. By applying the wave function assumptions $g(x_1,x_2)$ and boundary conditions, the wave functions in each region can be determined to establish the relationship between the incident and output state. The $g(x_1,x_2)$ in the darkly shaded areas represents the case where $x_2 < x_1$ and can be obtained through Bosonic statistics.}\label{fig:fig3}
\end{figure}

In the two-excitation space, the interacting eigenstate corresponding to the eigenvalue $E$ of the $e$-mode Hamiltonian in Eq.~(\ref{eq:eq3}) has following form
\begin{align}\label{eq:eq23}
|\phi^+\rangle = &\frac{1}{\sqrt{2}}\int dx_1 dx_2 g(x_1, x_2)C_e^\dagger(x_1) C_e^\dagger(x_2)|0, g\rangle  \nonumber\\
&+ \int dx \, e(x) C_e^\dagger(x) \sigma_+ |0, g\rangle,
\end{align}
where $g(x_1, x_2)$ represents the probability amplitude that there are two photons at the positions $x_1$ and $x_2$ of the waveguide and the giant atom is in its ground state $|g\rangle$. $e(x)$ represents the probability amplitude that there is one photon in the waveguide and the giant atom is in its excited state $|e\rangle$. By using eigenvalue equation $H_e |\phi^+\rangle = E |\phi^+\rangle$, we have
	
\begin{align}
&\left(i \sum_{i=1}^2\frac{\partial}{\partial x_i} + E \right) g  =V^{\prime}  \sum_{m=0}^1\Big\{  e(x_1) \delta(\overline{x}_{2m}) + e(x_2) \delta(\overline{x}_{1m})\Big\}, \label{eq:boundary_gx}\\
&\left(-i \frac{\partial}{\partial x} + \Omega - E \right) e(x) = -V^{\prime} \sum_{m=0}^1 \left[ g(x, \xi_{m}) + g(\xi_{m}, x) \right], \label{eq:boundary_ex}
\end{align}
where $g\equiv g(x_1,x_2)$, $V^{\prime}=V/(2\sqrt{2})$, $\delta(\overline{x}_{1m})\equiv \delta(x_{1}-\xi_m)$, and $\delta(\overline{x}_{2m})\equiv \delta(x_{2}-\xi_m)$.  The total energy $E$ of two photons satisfies the energy conservation condition $E = k + p$.
	
The photons are bosons, thus the solution of the two-photon wave function $g(x_1, x_2)$, corresponding to the two-photon state in the first  term of Eq.~(\ref{eq:eq23}),  should satisfy the exchange symmetric condition, i.e., $g(x_1, x_2) = g(x_2, x_1)$.  That is,  the  location exchange of two photons does not change the two-photon state. Thus, as shown in Fig.~\ref{fig:fig3}, we can use symmetry to obtain wave functions in full space by only studying the wave functions of two photons in the region $x_1 < x_2$.  Moreover, the giant atom is coupled to the waveguide via two coupling points, thus $g(x_1, x_2)$ is a piecewise continuous function,  it should satisfy the following boundary condition~\cite{Kaplan}
\begin{align} \label{eq:eq26}
g(\xi, x) &\equiv \frac{1}{2} \left[ g(\xi^-, x) + g(\xi^+, x) \right] \nonumber\\
&=g(x, \xi) \equiv \frac{1}{2} \left[ g(x, \xi^-) + g(x, \xi^+) \right],
\end{align}
with $\xi=0$ or $\xi=\xi_{1}$.

Our main purpose is to derive two-photon wave functions $g(x_1, x_2)$ corresponding to the two-photon states
\begin{equation}\label{eq:eq27}
\frac{1}{\sqrt{2}} \int dx_1 dx_2\, g(x_1, x_2) C_e^\dagger(x_1) C_e^\dagger(x_2) |0,g\rangle,
\end{equation}
in the first term of  Eq.~(\ref{eq:eq23}),  which includes solutions of the Bethe ansatz and bound states. To distinguish these two kinds of solutions represented in Eq.~(\ref{eq:eq27}), we use
\begin{equation}\label{F}
|\mathcal{F}_E \rangle=\frac{1}{\sqrt{2}} \int dx_1 dx_2 F(x_1, x_2) C_e^\dagger(x_1) C_e^\dagger(x_2) |0,g\rangle
\end{equation}
to denote the Bethe ansatz states, and
\begin{equation}\label{B}
|\mathcal{B}_E \rangle=\frac{1}{\sqrt{2}} \int dx_1 dx_2 B(x_1, x_2) C_e^\dagger(x_1) C_e^\dagger(x_2) |0,g\rangle
\end{equation}
 to denote the states that are not included in the solutions of Bethe ansatz. That is,  $g(x_1, x_2)\equiv F(x_1, x_2)$  corresponds to the state which can be obtained by Bethe ansatz method and $g(x_1, x_2)\equiv B(x_1, x_2)$ corresponds to the state which cannot be obtained by Bethe ansatz method. Here, the subscript $E$ denotes the total energy of two photons. In the following, we will derive these two kinds of states, and also $e(x)$ in Eq.~(\ref{eq:eq23}) is written as $e_F(x)$ and $e_B(x)$, which correspond to the two-photon states $|\mathcal{F}_{E}\rangle$  and $|\mathcal{B}_{E}\rangle$, respectively. Moreover, we use $|\phi^+_F\rangle$ to denote the interacting eigenstate with $g(x_1,x_2)=F(x_1,x_2)$ and $e(x)=e_F(x)$ and use $|\phi^+_B\rangle$ to denote the interacting eigenstate with $g(x_1,x_2)=B(x_1,x_2)$ and $e(x)=e_B(x)$ in Eq.~(\ref{eq:eq23}).

\subsection{Bethe ansatz solutions}\label{sec:twobethe}
	
We now apply the Bethe ansatz method to solve the two-photon wavefunctions and derive the eigenstates of the Hamiltonian $H_e$ in Eq.~(\ref{eq:eq3}). As shown in Fig.~\ref{fig:fig3}, two coupling points of the giant atom to the waveguide divide the real space into three different regions. Thus, the locations of  two photons have $6$ possible ways, which correspond to $6$ types of possible wave functions $F_{l,E}(x_1, x_2)$  in Eq.~(\ref{F}) with the formalism of the Bethe ansatz
\begin{align}\label{eq:eq30}
F_{l, E}(x_1, x_2) = A_l e^{ikx_1 + ipx_2} + B_l e^{ikx_2 + ipx_1},
\end{align}
with $l=1,\cdots,6$ for a given energy $E$ of two photons. Here, the subscript $l$ denotes the expression of $F(x_1, x_2)$ in Eq.~(\ref{F}) when two photons are in different regions.   For example, $F(x_1, x_2) \equiv{F}_{1,E}(x_1, x_2) $ denotes the wave function  when both photons locate at the left of the origin, i.e., $x_1<x_2<0$;   $F(x_1, x_2) \equiv{F}_{6,E}(x_1, x_2) $  denotes another wave function when two photons are in the right region of the position $\xi_{1}$,  i.e., $x_2>x_1>\xi_{1}$. The correspondence between $F_{l,E}(x_1, x_2) $ and photon locations is summarized in table~\ref{T1}. Hereafter, the two-photon state $|\mathcal{F}_{E}\rangle$ is written as  $|\mathcal{F}_{l,E}\rangle$ when we specify the wave function to $F_{l,E}(x_1, x_2)$.

\begin{table}[h]
\centering
\caption{The two-photon wave functions and corresponding regions that two photons locate.}\label{T1}
\begin{tabular}{|c|c|}
\hline
Two-photon wave functions  $F(x_1, x_2) $ & Photon locations \\ \hline
$F(x_1, x_2)\equiv F_{1,E}(x_1, x_2) $ & $x_{1}<x_{2}<0$  \\  \hline
$F(x_1, x_2)\equiv F_{2,E}(x_1, x_2) $ & $x_{1}<0$; \,\,$0<x_{2}<\xi_1$  \\ \hline
$F(x_1, x_2)\equiv F_{3,E}(x_1, x_2) $ & $x_{1}<0$;\,\,$\xi_1<x_2$  \\ \hline
$F(x_1, x_2)\equiv F_{4,E}(x_1, x_2) $ & $0<x_{1}<x_{2}<\xi_1$ \\ \hline
$F(x_1, x_2)\equiv F_{5,E}(x_1, x_2) $ & $0<x_{1}<\xi_1$; \,\,$\xi_1<x_{2}$  \\ \hline
$F(x_1, x_2)\equiv F_{6,E}(x_1, x_2) $ & $\xi_1<x_1<x_{2}$  \\ \hline
\end{tabular}
\end{table}

By applying the boundary conditions in Eq.~(\ref{eq:eq26}) to Eq.~(\ref{eq:boundary_gx}) and Eq.~(\ref{eq:boundary_ex}), we can determine the ratios $A_1/A_l$ and $B_1/B_l$ with $l=2,\cdots,6$ (details see Appendix~\ref{App:scattering}). Thus, it is clear that the coefficients $A_{6}$ and $B_{6}$ are determined by $A_{1}$ and $B_{1}$. To further obtain $A_{1}$ and $B_{1}$,  we can substitute $F_{l,E}(x_1,x_2)$ into the boundary conditions in Eq.~(\ref{eq:eq26}) to solve $e_F(x)$ via Eq.~(\ref{eq:boundary_ex}). As calculated in Appendix~\ref{App:scattering}, we have
\begin{align}
e_F(x) = \frac{V(A_1 + A_2 e^{ip\xi_1})}{\sqrt{2}(p - \Omega + \frac{iV^2}{4})} e^{ikx} + \frac{V(B_1 + B_2 e^{ik\xi_1})}{\sqrt{2}(k - \Omega + \frac{iV^2}{4})} e^{ipx},\label{eq:ex0}
\end{align}
for $x < 0$,
\begin{align}
e_F(x) = \frac{V(B_2 + A_4 e^{ip\xi_1})}{\sqrt{2}(p - \Omega + \frac{iV^2}{4})} e^{ikx} + \frac{V(A_2 + B_4 e^{ik\xi_1})}{\sqrt{2}(k - \Omega + \frac{iV^2}{4})} e^{ipx},
\end{align}
for $0 < x < \xi_1$, and
\begin{align}
e_F(x) = \frac{V(A_3 + A_5 e^{ik\xi_1})}{\sqrt{2}(k - \Omega + \frac{iV^2}{4})} e^{ipx} + \frac{V(B_3 + B_5 e^{ip\xi_1})}{\sqrt{2}(p - \Omega + \frac{iV^2}{4})} e^{ikx}, \label{eq:exxi}
\end{align}
for $x > \xi_1$.  Thus, we can determine the ratio $A_1/B_1$ by applying another boundary condition for $e_F(x)$ as
\begin{equation}
e_F(0^-) e^{i\beta_1} = e_F(\xi_1^+), \label{eq:exboundary}
\end{equation}
where $\beta_1$ represents the phase accumulation for a single photon propagating from left coupling point to right one~\cite{Cai2021PRA}. This boundary condition is valid only when $k,\, p\sim \Omega$, which can be considered as Markovian approximation. Substituting Eq.~(\ref{eq:ex0}) and Eq.~(\ref{eq:exxi}) into Eq.~(\ref{eq:exboundary}), we obtain
\begin{align}\label{eq:eq35}
\frac{A_1}{B_1} = \frac{k - p - i\Gamma_g}{k - p + i\Gamma_g},
\end{align}
where $\Gamma_g$ is the effective decay rate which has been defined in Eq.~(\ref{eq:gammag}). Here, we have used the relation $\exp(ipx)\approx\exp(ikx)\approx \exp(\beta_1)$ with $k \xi_1 = p \xi_1 = \Omega \xi_1 =\beta_1$ when Eq.~(\ref{eq:eq35}) is derived. Thus, all $F_{l,E}(x_1,x_2)$ can be obtained via the normalization. In our paper, we only discuss the Markovian case which can be solved analytically.
	
To conveniently describe the states obtained above, we now introduce two sets of complete bases $\{|S^{(e)}_{k,p}\rangle: k\leq p\}$  and $\{|A^{(e)}_{k,p}\rangle: k\leq p\}$ of the two-photon sates as
\begin{equation}
|S^{(e)}_{k,p}\rangle \equiv \frac{1}{\sqrt{2}} \int \mathrm{d}x_1 \mathrm{d}x_2 \, S_{k,p}  C_e^\dagger(x_1) C_e^\dagger(x_2) |0,g\rangle,\label{eq:Se}
\end{equation}
with
\begin{align}
\langle x_1, x_2 | S^{(e)}_{k,p} \rangle &= S_{k,p}(x_1, x_2) \nonumber\\
&= \frac{1}{2\sqrt{2}\pi}  \left(e^{ikx_1} e^{ipx_2} + e^{ikx_2} e^{ipx_1}\right),
\end{align}
and
\begin{equation}\label{eq:Ae}
|A^{(e)}_{k,p}\rangle \equiv \frac{1}{\sqrt{2}} \int \mathrm{d}x_1 \mathrm{d}x_2 \, A_{k,p}  C_e^\dagger(x_1) C_e^\dagger(x_2) |0,g\rangle,
\end{equation}
with
\begin{align}
\langle x_1, x_2 | A^{(e)}_{k,p} \rangle &= A_{k,p}(x_1, x_2) \nonumber\\
&= \text{sgn}(x) \frac{1}{2\sqrt{2}\pi} \left(e^{ikx_1} e^{ipx_2} - e^{ikx_2} e^{ipx_1}\right),
\end{align}
where we emphasize that $\{|S^{(e)}_{k,p}\rangle\}$  and $\{|A^{(e)}_{k,p}\rangle\}$ are complete for $k \leq p$~\cite{Schulz}. The distance between locations of two photons is $x = x_1 - x_2$  and $\text{sgn}(x) = \theta(x) - \theta(-x)$.  These two bases are both symmetric functions about $x_1$ and $x_2$, but they are not orthogonal to each other.
	
We construct the S-matrix following the process in single photon case. Since the interacting states have been obtained, we can obtain the in-state $|F_{i,E}(x_1,x_2)\rangle$ in the two-excitation case by replacing $|k^{(e)}\rangle$ and $|k^{(e)}_i\rangle$ in Eq.~(\ref{eq:LS1})  with $|\phi^+_F\rangle$ and $|F_{i,E}(x_1,x_2)\rangle$.  Similarly, the out-state $|F_{f,E}(x_1,x_2)\rangle$ can also be obtained via Eq.~(\ref{eq:LS2}). By using the solutions for $|\phi^+_F\rangle$  and the Lippmann-Schwinger equations in Eqs.~(\ref{eq:LS1}) and (\ref{eq:LS2}),  we find that  the  two-excitation wave function of the in-state $|F_{i,E}(x_1,x_2)\rangle$ and the out-state $|F_{f,E}(x_1,x_2)\rangle$ correspond to the states $|F_{1}(x_1,x_2)\rangle$ and $|F_{6}(x_1,x_2)\rangle$, respectively. They can be written as
\begin{align}
F_{i,E}(x_1,x_2)&=F_{1,E}(x_1,x_2)\theta(x_2-x_1)\nonumber\\
                          &+F_{1,E}(x_2,x_1)\theta(x_1-x_2),\label{eq:41}\\
F_{f,E}(x_1,x_2)&=F_{6,E}(x_1,x_2)\theta(x_2-x_1)\nonumber\\
                         &+F_{6,E}(x_2,x_1)\theta(x_1-x_2),\label{eq:42}
\end{align}
with $\theta(x)=1$ for $x>0$ and  $\theta(x)=0$ for $x<0$. That is, Eqs.~(\ref{eq:41}) and (\ref{eq:42}) represent in-states and out-states for all possible locations of two photons, not only are limited to the case $x_1<x_2$.

The in-states  $|F_{i,E}(x_1,x_2)\rangle$ can be conveniently expressed in two complete bases in Eq.~(\ref{eq:Se}) and Eq.~(\ref{eq:Ae}) as,
\begin{equation}
|F_{i,E}(x_1,x_2) \rangle = \frac{1}{\sqrt{4\Delta^2 + \Gamma_g^2}} [2\Delta |S^{(e)}_{k,p}\rangle + i \Gamma_g |A^{(e)}_{k,p}\rangle], \label{sca}
\end{equation}
with $k\leq p$, where $2\Delta = k - p$ characterizes  the energy difference between two photons. We also find that the out-state $|F_{f,E}(x_1,x_2)\rangle$ satisfies the relation $|F_{f,E}(x_1,x_2)\rangle=t_{k}t_{p}|F_{i,E}(x_1, x_2)\rangle$, where $t_{k}$ in the Markovian approximation is given by
\begin{equation}
t_k = \frac{2(k - \Omega) - i\Gamma_e^*}{2(k - \Omega) + i\Gamma_e},\label{eq:tkM}
\end{equation}
with
\begin{equation}
\Gamma_e = \frac{V^2 (1 + e^{i\beta_1})}{2},\label{eq:Gamma_e}
\end{equation}
where $\Gamma_e^{*}$ is the complex conjugate of $\Gamma_e$ and $t_{p}$ is obtained by replacing $k$ by $p$ in Eq.~(\ref{eq:tkM}). Then, we can construct the S-matrix for two-photon scattering of the $e$-mode with the Bethe ansatz solutions as
\begin{align}\label{eq:S_F}
S_{F}=\sum_{k \leq p}|F_{f,E}\rangle\langle F_{i,E}|=\sum_{k \leq p}t_kt_p|F_{i,E}\rangle\langle F_{i,E}|.
\end{align}
It is clear that the states $|F_{i,E}(x_1, x_2)\rangle$ are the eigenstates of the S-matrix with eigenvalue $t_kt_p$, but they are incomplete and cannot span the whole two-photon Hilbert space governed by $H_0$, so we need to subtract the states $|F_{i,E}(x_1, x_2)\rangle$ from the entire space $\{ |S^{(e)}_{k,p} \rangle: k \leq p \}$ to obtain other in-states.

\subsection{Bound states}\label{sec:twobound}
	
We have found in-states and out-states corresponding to the Bethe ansatz solutions. However, to construct the entire S-matrix, we need to find other in-states and out-states corresponding to the two-photon bound states~\cite{Shen2007PRA}.  Under the Markovian approximation, as shown in Appendix~\ref{App:bound},  we subtract the in-states $|F_{i,E}(x_1, x_2)\rangle$ from the entire space,  thus we can obtain other possible in-states $|B_{i,E}\rangle$ of two photons
\begin{equation}
\left|B_{i,E}\right\rangle \equiv \frac{1}{\sqrt{2}}\int dx_1 dx_2 \, B_{i,E}(x_1, x_2)  C_e^\dagger(x_1) C_e^\dagger(x_2) |0, g\rangle,
\label{eq:bo}
\end{equation}
which are not included in those in-states $|F_{i,E}(x_{1},x_2)\rangle$ in Eq.~(\ref{sca}) derived from the Bethe ansatz solutions.  Similar to the states $|F_{i,E}(x_1,x_2)\rangle$,  the subscripts $i$ and $E$ denote  the in-state $|B_{i,E}\rangle $ with the total energy $E$ of two photons. The wave function $B_{i,E}(x_1, x_2)$ in Eq.~(\ref{eq:bo}) is expressed as
\begin{equation}
B_{i,E}(x_1, x_2) =\langle x_1, x_2 | B_{i,E} \rangle = \sqrt{\frac{\Gamma_g}{4\pi}} e^{iEx_c - \Gamma_g |x|/2},\label{eq:B(x_1,x_2)}
\end{equation}
which satisfies the normalization condition
\begin{equation}
\langle B_{i,E'} | B_{i,E} \rangle = \delta(E - E').
\end{equation}
Here, $x_c = (x_1 + x_2)/2$ and $x = x_1 - x_2$. It can be seen that the center of the two-photon wave function for the bound state propagates with the energy $E$, but the wave function decays at a rate $ \Gamma_g $ when the relative coordinate $x$ is increased. We note that two-photon bound states $\left|B_{i,E}\right\rangle$ are orthogonal to the two-photon states $\left|F_{i,E}\right\rangle$.

To obtain the scattering matrix corresponding to the bound states, we need to obtain both the in-states and out-states. Similar to the two-photon wave function with Bethe ansatz solutions, we write out the two-photon wave function $B(x_1,x_2)$ in Eq.~(\ref{B})  for the bound states in six different regions of two photons. According to the expression of the in-states in Eq.~(\ref{eq:bo}), we make another ansatz that the two-photon wave functions  $B(x_1,x_2)$  in different regions have following form
\begin{align}
 B_{l,E}(x_1, x_2) =t_{l}\sqrt{\frac{\Gamma_g}{4\pi}} e^{(iE + \Gamma_g)x_1/2} e^{(iE - \Gamma_g)x_2/2}, \label{eq:bound}
\end{align}
with $l=1,\,\cdots, 6$. Here,  the correspondence between $B_{l,E}(x_1,x_2)$ and photon location can be obtained by replacing $F_{l,E}$ with $B_{l,E}$ in the table~\ref{T1}.  We use the $B_{l,E}(x_1,x_2)$ to denote the  wave function corresponding to the $l$th location of two photons. For example, $B(x_1,x_2)=B_{1,E}(x_1,x_2)$ is the wave function when both photons are in the left of the origin, i.e., $x_1<x_2<0$; $B(x_1,x_2)=B_{6,E}(x_1,x_2)$ is the wave function when both photons are in the right of the position $\xi_1$, i.e., $x_2>x_1>\xi_1$.   The coefficients $t_l$ for the wave function $B_{l,E}(x_1,x_2)$  with $l=2,\cdots,6$ can be obtained  via the ratio $t_l/t_1$  by applying the boundary conditions in Eq.~(\ref{eq:eq26}) (see Appendix~\ref{App:bound}). By using the Lippmann-Schwinger equations, it can be proved that the wave functions corresponding to the state $|B_{1,E}(x_1,x_2)\rangle $ and and the in-state $ |B_{i,E}(x_1, x_2)\rangle$ in Eq.~(\ref{eq:B(x_1,x_2)}) have the following relation
\begin{align}
B_{i,E}(x_1,x_2)&=B_{1,E}(x_1,x_2)\theta(x_2-x_1)\nonumber\\
&+B_{1,E}(x_2,x_1)\theta(x_1-x_2),
\end{align}
thus we have $t_{1}=1$.	

To verify if the boundary condition in Eq.~(\ref{eq:exboundary}) can still be applied to the wave function $e_B(x)$,  we substitute $B_{l,E}(x_1,x_2)$ into the boundary conditions in Eq.~(\ref{eq:boundary_ex}) and Eq.~(\ref{eq:eq26}) (see Appendix~\ref{App:bound}), and calculate the ratio
\begin{align}\label{eq:eq49}
\frac{e_{B}(0^{-})}{e_{B}(\xi_1^{+})} &= \frac{1 + e^{(-iE + \Gamma_g)\xi_1/2}}{1 + e^{(iE + \Gamma_g)\xi_1/2}} \nonumber \\
& \times \frac{2E - 2i\Gamma_g - 4\Omega + iV^2 (1 + e^{(iE + \Gamma_g)\xi_1/2})}{2E + 2i\Gamma_g - 4\Omega - iV^2 (1 + e^{(-iE + \Gamma_g)\xi_1/2})}.
\end{align}
We find that  Eq.~(\ref{eq:eq49}) can be simplified to
\begin{equation}
e_{B}(0^{-}) e^{i\beta_1} = e_{B}(\xi_1^{+}).
\end{equation}
when $\Gamma_g \xi_1 \approx 0$, which means that the relaxation time $1/\Gamma_{g}$ of the giant atom is much longer than the travelling time $\xi_{1}/v_{g}$ of the photons through two coupling points. This is consistent with the assumption $\exp(ip\xi_1)\approx\exp(ik\xi_1)\approx \exp(\beta_1)$ for $e_F(x)$, which results in the relation $\exp(iE/2\xi_1)=\exp(i(k+p)/2\xi_1)\approx \exp(\beta_1)$ in the amplitude $e_B(x)$.
	
Under the Markovian approximation with $\Gamma_g \xi_1 \approx 0$ and $\exp(ip\xi_1)\approx\exp(ik\xi_1)\approx \exp(\beta_1)$, we can obtain the transmission coefficient $t_B$ via the two-photon states $|\mathcal{B}_E\rangle$ in Eq.~(\ref{B})  corresponding to the Hamiltonian $H_e$ as
\begin{align}
t_B &=t_6= \frac{E - 2\Omega - i\Gamma_g - i\Gamma_e^*}{E - 2\Omega + i\Gamma_g + i\Gamma_e}.
\end{align}
By using  the Lippmann-Schwinger equations, we can also obtain  the wave functions $B_{f,E}(x_1,x_2)$ for the out-states $|B_{f,E}(x_1,x_2)\rangle$ via interacting eigenstate $|\phi^+_B\rangle$. Then, the wave functions of out-states $B_{f,E}(x_1,x_2)$ for the $e$-mode can be written as
\begin{align}
B_{f,E}(x_1,x_2)&=B_{6,E}(x_1,x_2)\theta(x_2-x_1)\nonumber\\
                           &+B_{6,E}(x_2,x_1)\theta(x_1-x_2),
\end{align}
Since the wave function $B_{f,E}(x_1,x_2)$ can be also expressed as $t_BB_{i,E}(x_1,x_2)$, the S-matrix for the bound states corresponding to the $e$-mode can be written as
\begin{align}
S_B=\sum_E|B_{f,E}\rangle\langle B_{i,E}|=\sum_Et_B|B_{i,E}\rangle\langle B_{i,E}|.\label{eq:S_B}
\end{align}

It is clear that the bound state is the eigenstate of the S-matrix in Eq.~(\ref{eq:S_B})  with the eigenvalue $t_B$. The set $\{|F_{i,E}\rangle : \forall k \leq p\} \cup \{|B_{i,E}\rangle\}$ forms a complete basis of  the two-photon Hilbert space corresponding to the Hamiltonian in Eq.~(\ref{eq:eq3}), allowing any symmetric function of $x_1$ and $x_2$ of the in-state to be expanded using $\{F_{i,E}(x_1, x_2), B_{i,E}(x_1, x_2)\}$. Since we have determined all the eigenstates and eigenvalues of the S-matrix in the Markovian regime, the S-matrix $ S_{ee} $ for the two-photon scattering of the $e$-mode is given by:
\begin{equation}\label{eq:eq54}
S_{ee} \equiv S_F+S_B.
\end{equation}
Here, $S_F$ is the S-matrix given in Eq.~(\ref{eq:S_F}) and $S_B$ is the S-matrix given in Eq.~(\ref{eq:S_B}). Thus, the total S-matrix of the two-photon scattering for both the $e$-mode and $o$-mode is:
\begin{equation}
S = S_{ee} + S_{oo}.\label{eq:S}
\end{equation}
Here, $ S_{oo} $ is the identity matrix in the $o$-mode space and describes the free propagation of photons in the $o$-mode space with
\begin{equation}\label{eq:60}
S_{oo} = \sum |S^{(o)}_{kp}\rangle \langle S^{(o)}_{kp}|,
\end{equation}
where $|S^{(o)}_{kp}\rangle$ is a two-photon state in the $o$-mode space with the momentums $k$ and $p$ and is given by replacing the label $e$ with $o$ in Eq~(\ref{eq:Se}). Thus, the total S-matrix in Eq.~(\ref{eq:S}) under the Markovian approximation incorporates contributions from both the $e$-mode and $o$-mode. The $e$-mode accounts for the interaction effect with the giant atom, while the $o$-mode represents the free propagation of photons.

 In summary, we have derived the two-photon scattering matrix in Eq.~(\ref{eq:S}) including the $e$-mode and $o$-mode. That is, the output state for arbitrarily incident two-photon state of the $e$-mode and $o$-mode can be obtained via the scattering matrix in Eq.~(\ref{eq:S}). The two-photon scattering matrix  for the  $L$-mode and $R$-mode can be directly obtained via the relation of the $e$-mode and $o$-mode with the $L$-mode and $R$-mode in Eq.~(\ref{eq:dec_eo}). Thus, the output state for arbitrarily incident two-photon state of the $L$-mode and $R$-mode can be obtained via the scattering matrix in Eq.~(\ref{eq:S}) and transform in Eq.~(\ref{eq:dec_eo}).

\section{Two-photon scattering for given incident two-photon states}\label{sec:two mode}

We have obtained the two-photon S-matrix in Eq.~(\ref{eq:S}). Thus, we can analyze the properties of the  two-photon scattering for any incident two-photon state with the momenta $k_1$ and $p_1$. For concreteness, we here mainly use an incident two-photon state in the $R$-mode space as an example to study the two-photon scattering. However, the S-matrix given in Eq.~(\ref{eq:S}) is derived in the $o$-mode and $e$-mode spaces, thus below we first show how to calculate the output state corresponding to an incident state in the $e$-mode space, then we further study the two-photon scattering
for an incident two-photon state in the $R$-mode space.

\subsection{Scattering for incident two-photon states of $e$-mode}

In this section, we study the two-photon scattering when the incident two-photon state is in the $e$-mode space. To compare the results with those of two-photon scattering by a natural atom~\cite{Shen2007PRA}, as an example, we take the two-photon state in Eq.~(\ref{eq:Se}) with $k=k_1$ and $p=p_1$  in the $e$-mode space as an incident state, i.e.,  $| X^{(e)}_{\text{in}} \rangle=|S^{(e)}_{k_1,p_1}\rangle$, which is the same as that in Ref.~\cite{Shen2007PRA}. Here,  the subscript ``in''  denotes incident two-photon states.  In this case,  the output state $| X^{(e)}_{\text{out}}\rangle$  is still in the $e$-mode space and can be obtained by applying the S-matrix for the $e$-mode  in Eq.~(\ref{eq:eq54})  on the incident state  $| X^{(e)}_{\text{in}} \rangle=|S^{(e)}_{k_1,p_1}\rangle$, i.e.,
\begin{equation}\label{eq:61}
| X^{(e)}_{\text{out}} \rangle=S_{ee}|S^{(e)}_{k_{1},p_{1}}\rangle\equiv \left[S_{F}+S_B\right]|S^{(e)}_{k_{1},p_{1}}\rangle.
\end{equation}

If we assume that the total energy of two incident photons with the momentums $k_1$ and $p_1$ is $E_{1}$, then the corresponding elements of the scattering matrix for the output two-photon state with the momentums $k_{2}$ and $p_{2}$ in the $e$ space can be given as (see Appendix~\ref{App:B})
\begin{align}
&\langle S^{(e)}_{k_{2},p_{2}}|S_{ee}|S^{(e)}_{k_{1},p_{1}}\rangle=t_{k_{1}}t_{p_{1}}\delta(k_{1}-k_{2}) \delta(p_{1}-p_{2})\nonumber\\&+t_{k_{1}}t_{p_{1}}\delta(k_{1}-p_{2}) \delta(k_{2}-p_{1})+B\delta(E_{1}-E_{2}), \label{eq:Bk}
\end{align}
where
\begin{align}
B=\frac{16i\Gamma_{g}^2\Sigma}{\pi[4\Delta_1^2-\Sigma^2][4\Delta_2^2-\Sigma^2]}, \label{eq:Bk1}
\end{align}
with $ \Sigma=E_1-2\Omega+i\Gamma_{e}$. Here, $E_{2}$ is total energies of the output two photons. $ 2\Delta_1 = k_1 - p_1 $ ($ 2\Delta_2 = k_2 - p_2 $) characterizes the energy difference between two photons in the incident (output) state. The first two delta functions in Eq.~(\ref{eq:Bk}) represent the properties of the scattering state, where the photons either retain their original momenta or exchange momenta. The final term $B$ in Eq.~(\ref{eq:Bk}) describes the properties of the bound state, where the energy conservation $ E_1 = E_2 $ is applied and there are different momentum distributions compared to the incident state.

We can also derive the real-space representation of the output state
\begin{align}\label{eq:eq60}
&\langle \{x_{c}, x \}^{(e)}| X^{(e)}_{\text{out}} \rangle = \langle \{x_{c}, x \}^{(e)} | S_{ee} | X^{(e)}_{\text{in}} \rangle \nonumber \\
&= e^{i E_{1} x_{c}} \frac{\sqrt{2}}{2 \pi} \biggl[ t_{k_{1}} t_{p_{1}} \cos(\Delta_{1} x) - \frac{4 \Gamma_{g}^2}{4 \Delta_1^2 - (E_1 - 2 \Omega + i \Gamma_{e})^2} \nonumber \\
& \quad \times e^{i (E_1 - 2 \Omega) |x| / 2 - \Gamma_{e} |x| / 2} \biggr],
\end{align}
for the incident two-photon state $| X^{(e)}_{\text{in}} \rangle=|S^{(e)}_{k_1,p_1}\rangle$.  Comparing the results in Eqs.~(\ref{eq:Bk}) and (\ref{eq:eq60}) with those of the two-photon scattering by the natural atom~\cite{Shen2007PRA}, we find that they have the similar expressions, but here the decay rates $\Gamma_e$ and $\Gamma_g$ are closely related to the photon interference effect between two coupling points. Such effect cannot be found in the two-photon scattering by the natural atom. We here emphasize that the output states are the same as the incident states when the incident two-photon states are limited to the $o$-mode,  because the $o$-mode is decoupled from the giant atom.

\subsection{Scattering for incident two-photon states of $R$-mode }

Let us now study the two-photon scattering when the incident two-photon state is in the $R$-mode space. As an example, we assume that the incident state in the $R$-mode space is written as
\begin{align}\label{eq:eq61}
\left|X^{(R)}_{\mathrm{in}}\right\rangle & \equiv|S^{(R)}_{k_{1},p_{1}}\rangle=\frac{1}{4\pi}\int dx_{1}dx_{2}\left(e^{ik_{1}x_{1}+ip_{1}x_{2}} \right.\nonumber\\
&\left.+e^{ik_1x_2+ip_1x_1}\right)C_R^\dagger(x_1)C_R^\dagger(x_2)|0,g\rangle,
\end{align}
where $ k_1 $ and $ p_1 $ are the momentums of two incident photons, and $ x_1 $ and $ x_2 $ are their locations. Considering that the S-matrix is given in the $e$-mode and $o$-mode spaces, we first decompose $C_R^\dagger $ into $ C_e^\dagger $ and $ C_o^\dagger$ using the inverse of Eq.~(\ref{eq:dec_eo}). We then apply the S-matrix given in Eq.~(\ref{eq:S}) to Eq.~(\ref{eq:eq61}) and obtain the output state in the $e$-mode and $o$-mode spaces as for Eq.~(\ref{eq:61}). Finally, using the relation given in Eq.~(\ref{eq:dec_eo}), the output state $ \left| X_{\text{out}} \right\rangle$ in the $R$-mode and $L$-mode spaces corresponding to the incident state in Eq.~(\ref{eq:eq61}) can be obtained as
\begin{align}
\left|X_{\mathrm{out}}\right\rangle & =S|X^{(R)}_{\mathrm{in}}\rangle  \nonumber\\
&=\frac{1}{\sqrt{2}}\int dx_{1}dx_{2}t_{2}(x_{1},x_{2})C_{R}^{\dagger}(x_{1})C_{R}^{\dagger}(x_{2})|0,g \rangle \nonumber\\
&+\frac{1}{\sqrt{2}}\int dx_1dx_2r_2(x_1,x_2)C_L^\dagger(x_1)C_L^\dagger(x_2)|0,g\rangle \nonumber\\
&+\int dx_1dx_2r_t(x_1,x_2)C_R^\dagger(x_1)C_L^\dagger(x_2)|0,g \rangle,
\label{eq:out}
\end{align}
with
\begin{align}
&t_2(x_1,x_2) =e^{iE_{1}x_{c}}\frac{\sqrt{2}}{2\pi}\biggl[\overline{t}_{k_{1}}\overline{t}_{p_{1}}\cos(\Delta_{1}x) \nonumber\\
&-\frac{\Gamma_{g}^2}{4\Delta_1^2-(E_1-2\Omega+i\Gamma_{e})^2}e^{i(E_1-2\Omega)|x|/2-\Gamma_{e}|x|/2}\biggr], \label{trans:t2} \\
&r_2(x_1,x_2) =e^{-iE_{1}x_{c}}\frac{\sqrt{2}}{2\pi}\biggl[\overline{r}_{k_{1}}\overline{r}_{p_{1}}\cos(\Delta_{1}x) \nonumber\\
&-\frac{\Gamma_{g}^2}{4\Delta_1^2-(E_1-2\Omega+i\Gamma_{e})^2}e^{i(E_1-2\Omega)|x|/2-\Gamma_{e}|x|/2}\biggr], \label{trans:t3}
\end{align}
and
\begin{align}		r_t(x_{1},x_{2})&=\frac{1}{2\pi}e^{i(E_{1}/2)x}\biggl[\overline{t}_{k_{1}}\overline{r}_{p_{1}}e^{2i\Delta_{1}x_{c}}+\overline{r}_{k_{1}}\overline{t}_{p_{1}}e^{-2i\Delta_{1}x_{c}}\nonumber\\ &-\frac{2\Gamma_{g}^{2}}{4\Delta_{1}^{2}-(E_{1}-2\Omega+i\Gamma_{e})^{2}}e^{i(E_{1}-2\Omega)|x_{c}|-\Gamma_{e}|x_{c}|}\biggr], \label{trans:t4}
\end{align}
with the total energy $E_{1}$ of two incident photons, where $ \overline{t}_{k_{1}}(\overline{t}_{p_{1}})$ and $ \overline{r}_{k_{1}} (\overline{r}_{p_{1}})$ are the transmission and reflection coefficients in the $ R $ and $L$ spaces which can be obtained from Eq.~(\ref{eq:tk}) by replacing $\beta_1^{\prime}$ with $\beta_1$ for Markovian approximation. Here, $t_2(x_{1},x_{2})$ is the probability amplitude for both photons to propagate toward  the right. Similarly, $r_2(x_{1},x_{2})$ is the probability amplitude for both photons to propagate toward the left,  and $ r_t (x_{1},x_{2})$ is the probability amplitude that one photon propagates toward the left and the other one propagates toward the right. Our result here agrees well with the those calculated by using Lippmann-Schwinger (LS) method~\cite{Gu2023PRA}. We find that the first terms in the coefficients $t_2(x_{1},x_{2})$, $r_2(x_{1},x_{2})$ and $r_t(x_{1},x_{2})$ contribute to the scattering state, however the second terms contribute to the bound state.
	
We now discuss the effects of $\delta E_1 = E_1 - 2\Omega$ and $2\Delta_1=k_1-p_1$ on the coefficients $t_2(x_{1},x_{2})$, $r_2(x_{1},x_{2})$ and $r_t(x_{1},x_{2})$.  It is clear that $\delta E_1$ is difference between the total energy of two incident photons and the double transition energies of the giant atom. We first consider the case that two incident photons have the same energy, i.e., $k_1=p_1$. Moreover, as studied in Sec~\ref{sec:single}, we assume that each incident single photon satisfies the resonant scattering condition $k_1=p_1=\Omega+\Delta_L$ with $\Delta_L=\Gamma \sin(\beta_1)/4$  given in Eq.~(\ref{eq:gammag}),  then the energy of the two photons is $E_{1}=2\Omega+2\Delta_L$ for the resonant scattering, and the detuning $\delta E_1$ can be expressed by $\delta E_1=E_{1}-2\Omega=2\Delta_L$. In this case, the coefficients $t_2(x_{1},x_{2})$, $r_2(x_{1},x_{2})$ and $r_t(x_{1},x_{2})$ in Eqs.~(\ref{trans:t2}), (\ref{trans:t3}) and (\ref{trans:t4})  are simplified to
\begin{align}
t_{2} (x_{1},x_{2})&= \frac{e^{iE_1 x_{c}}}{\sqrt{2}\pi}\left[\frac{\Gamma_{g}^2\times e^{ i\Delta_L|x| - \Gamma_{e} |x| / 2}}{\left(2\Delta_L + i \Gamma_{e}\right)^2}\right], \label{eq:t2} \\
r_{2} (x_{1},x_{2})&= \frac{e^{-iE_1 x_{c}}}{\sqrt{2}\pi}\left[1 + \frac{\Gamma_{g}^2\times e^{i \Delta_L|x| - \Gamma_{e} |x| / 2}}{\left(2\Delta_L + i \Gamma_{e}\right)^2} \right], \label{eq:r2}\\
r_t (x_{1},x_{2})&= \frac{e^{iE_1 x / 2}}{2\pi}\left[\frac{2 \Gamma_{g}^2\times e^{i \Delta_L |x_{c}| - \Gamma_{e} |x_{c}|}}{\left(2\Delta_L + i \Gamma_{e}\right)^2} \right].\label{eq:rt}
\end{align}
It is clear that the maximum values of the coefficients $t_2(x_{1},x_{2})$, $r_2(x_{1},x_{2})$ and $r_t(x_{1},x_{2})$	do not correspond to $E_{1}=2\Omega$ as for the case of the natural atom, but are adjusted by $\Delta_L$ determined by the distance between two coupling points.

\begin{figure}
\centering		
\includegraphics[width=1\linewidth]{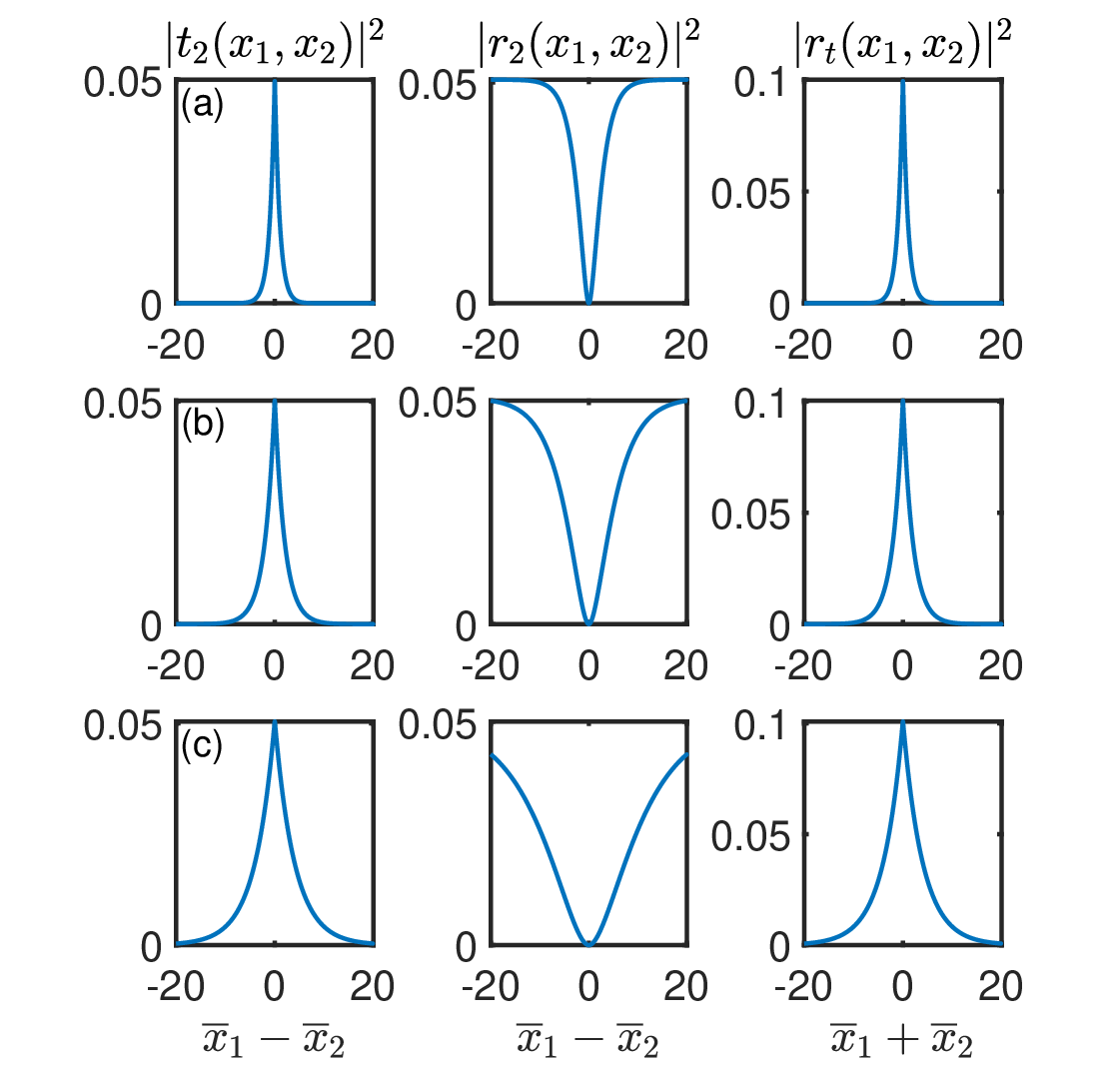}
\caption{Spatial variations of wave functions $|t_2(x_1,x_2)|^2$, $|r_2(x_1,x_2)|^2$, and $|r_t(x_1,x_2)|^2$ of the output state  with $k_1=p_1=\Omega+\Delta_L$  for $\beta_1=k_1\xi_1 = 0$ in the first row (a), $\pi/2$  for the second row in (b), and $2\pi/3$  for the third row in (c), corresponding to $\Delta_L=0$, $\Delta_L=\Gamma/4$ and $\Delta_L=\sqrt{3}\Gamma/8$, from top row to bottom bow. Other  parameters are chosen  as $\Delta_1 = k_1 - p_1= 0$, $\delta E_1 = E_1 - 2\Omega = 2\Delta_L$, and $\overline{x}_i = x_i \Gamma$.}
\label{fig:fig4}
\end{figure}
	
\begin{figure}
\centering
\includegraphics[width=1\linewidth]{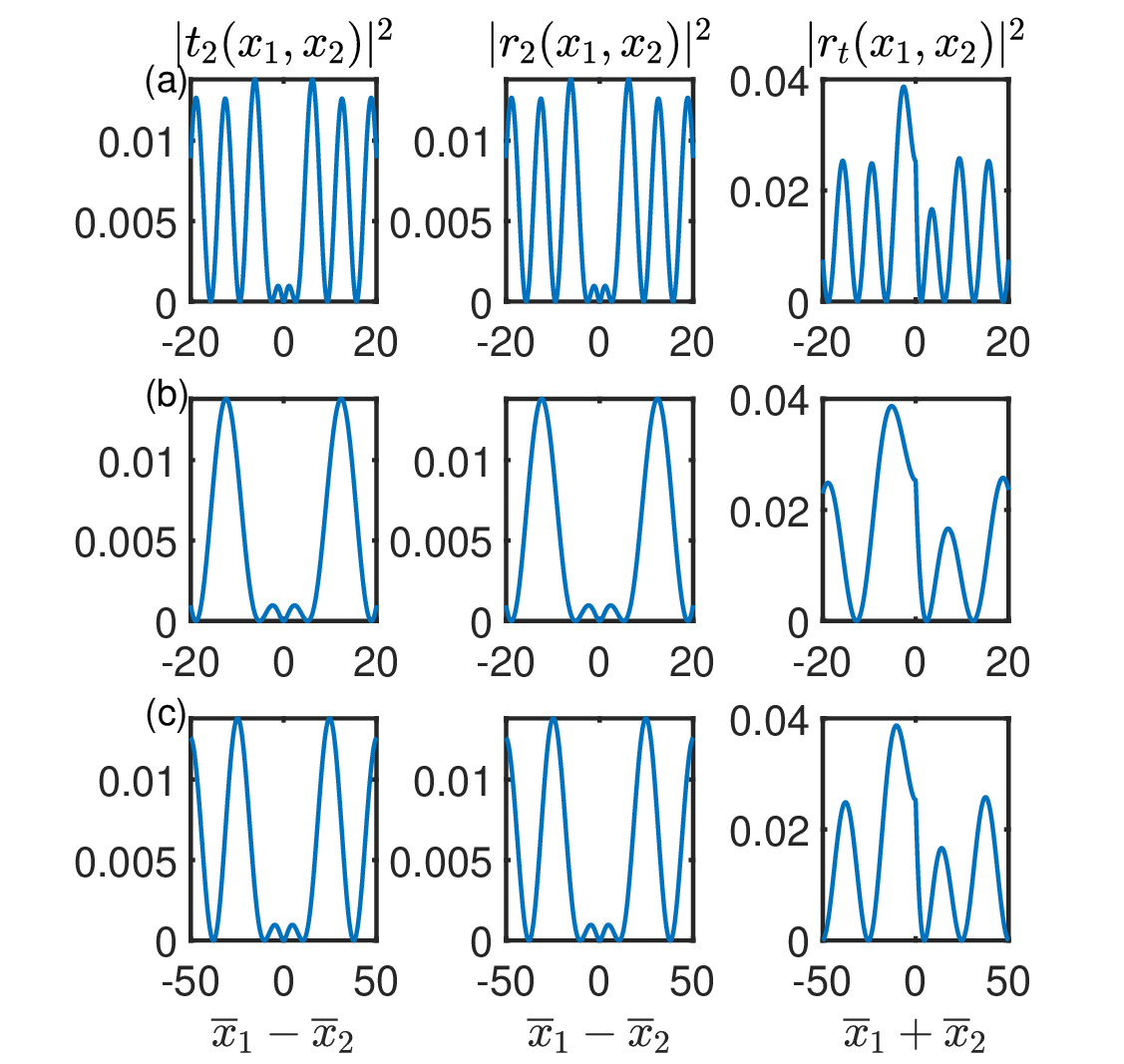}
\caption{Spatial variations of the wave functions $|t_2(x_1,x_2)|^2$, $|r_2(x_1,x_2)|^2$, and $|r_t(x_1,x_2)|^2$ of the output state for $\beta_1=k_1\xi_1 = 0$  in the first row (a), $\pi/2$  in the second row (b), and $2\pi/3$  in the third row (c), corresponding to $\Delta_L=0$, $\Delta_L=\Gamma/4$ and $\Delta_L=\sqrt{3}\Gamma/8$, from top row to bottom bow. Other parameters are chosen as $\Delta_1 = (k_1 - p_1)/2 = -\Gamma(1 + \cos(\beta_1))/4$, $\delta E_1 = E_1 - 2\Omega = 2\Delta_L$, and $\overline{x}_i = x_i \Gamma$.}
\label{fig:fig5}
\end{figure}
	
\begin{figure}
\centering
\includegraphics[width=1\linewidth]{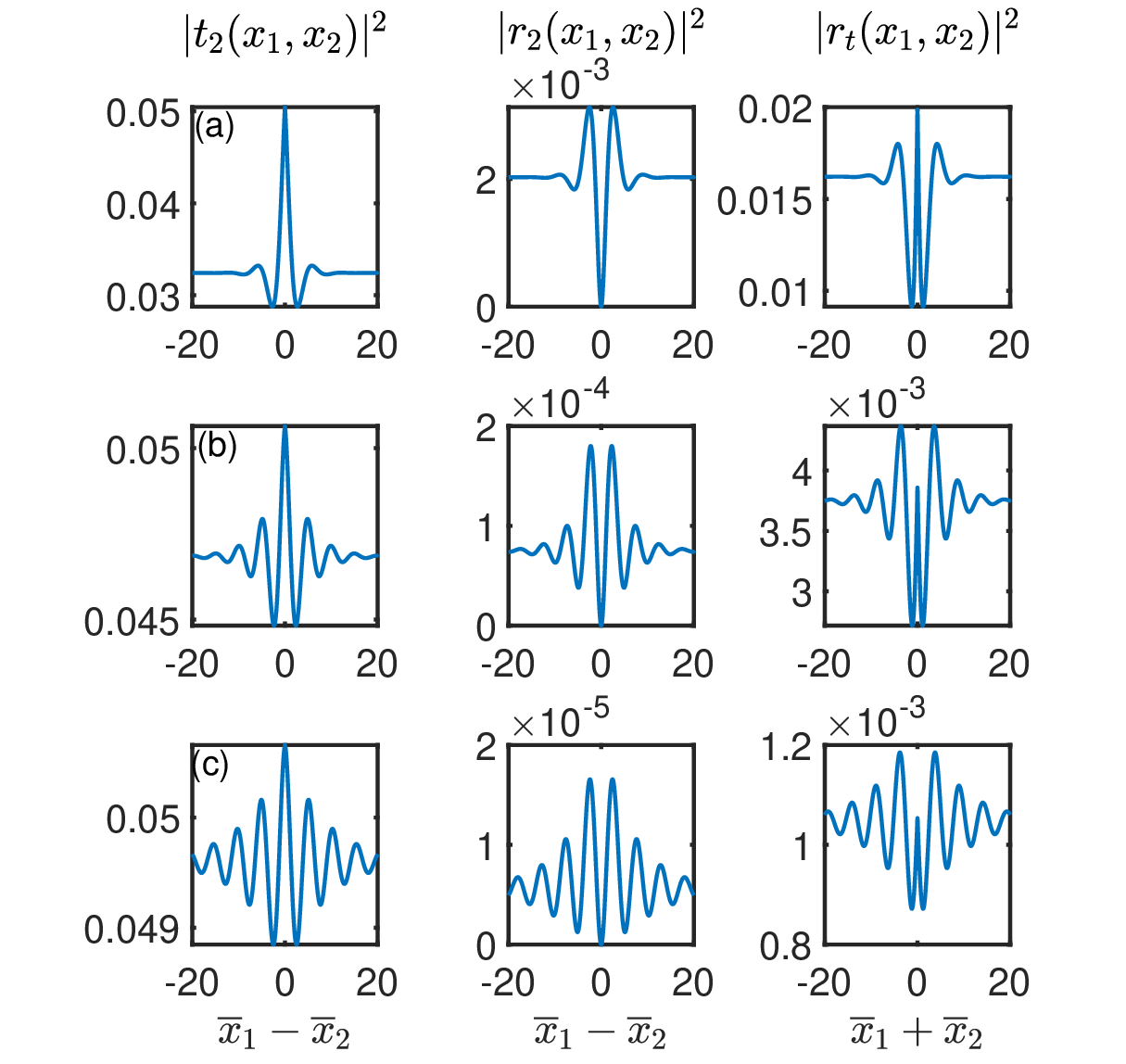}
\caption{Spatial variations of the wave functions $|t_2(x_1,x_2)|^2$, $|r_2(x_1,x_2)|^2$, and $|r_t(x_1,x_2)|^2$ of the output state for $\beta_1=k_1\xi_1 = 0$ in the first row (a), $\pi/2$ in the second row (b), and $2\pi/3$ in the first row (c), corresponding to $\Delta_L=0$, $\Delta_L=\Gamma/4$ and $\Delta_L=\sqrt{3}\Gamma/8$, from top row to bottom bow. Other parameters are chosen as $\Delta_1 = (k_1 - p_1) = 0$, $\delta E_1 = E_1 - 2\Omega = -2\Gamma$, and $\overline{x}_i = x_i \Gamma$.}
\label{fig:fig6}
\end{figure}
	
\begin{figure*}
\centering
\includegraphics[width=1\linewidth]{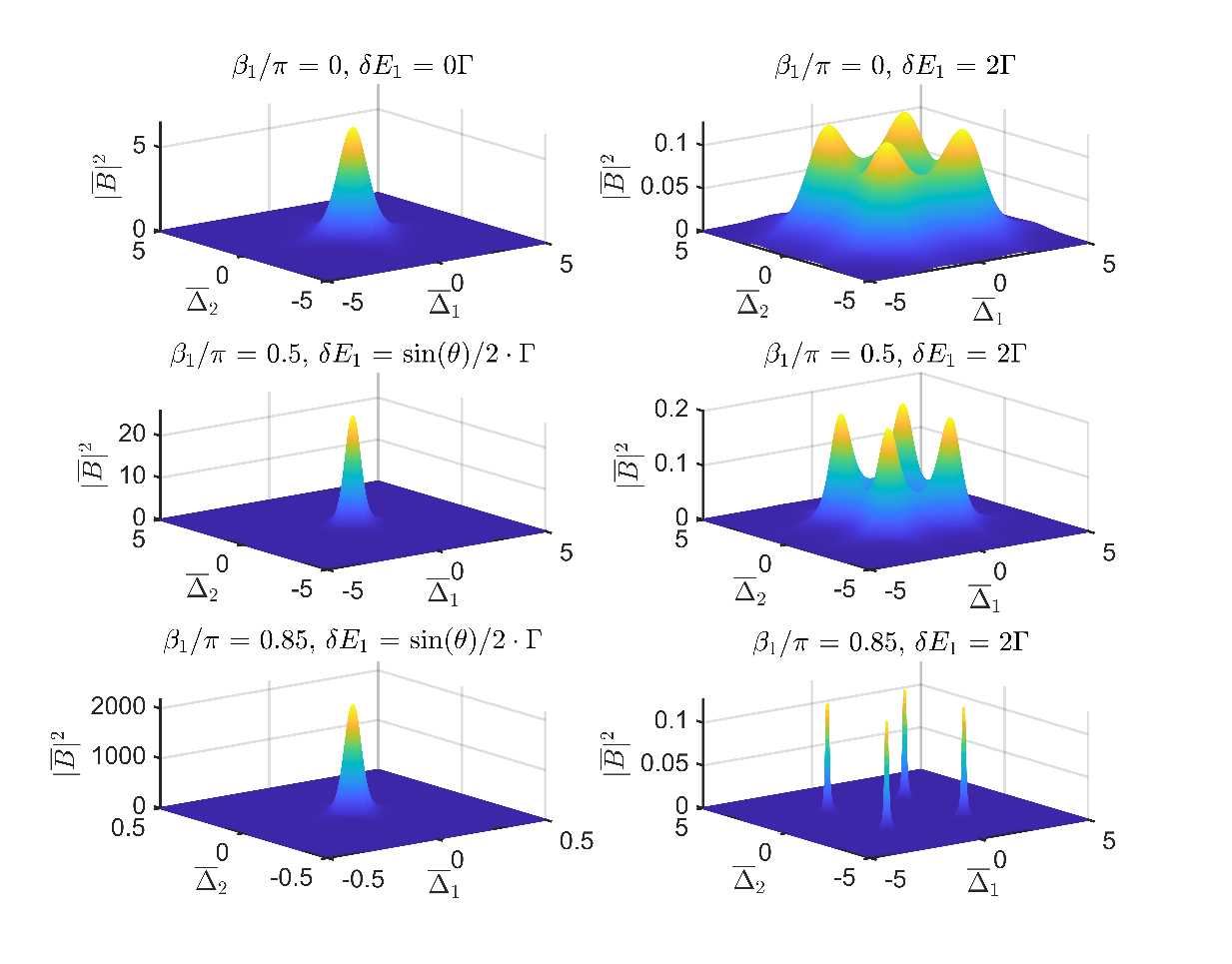}
\caption{Bound state distribution in momentum space as a function of $\overline{\Delta}_1$ and $\overline{\Delta}_2$ at various energy and phase. From top row to bottom row, the phase $\beta_1=k_0\xi_1$ are $0$, $0.5\pi$, and $0.85\pi$. The left column represents the case where each of two incident photons is resonant with the giant atom for each phase. The right column illustrates a scenario with fixed incident energy but not satisfies single photon resonant conditions for each incident photon. The other parameters are $\overline{\Delta}_{1(2)}= \Delta_{1(2)}/(\Gamma/2)$, $\overline{B} = (\Gamma/2)B$, and $\delta E_1 = (E_1 - 2\Omega)$, $\Gamma=V^2$.}
\label{fig:fig7}
\end{figure*}
	
In Fig.~\ref{fig:fig4}, we plot the spatial distribution $|t_2(x_{1},x_{2})|^2$, $|r_2(x_{1},x_{2})|^2$ and $|r_t(x_{1},x_{2})|^2$ of the wave function  given in Eqs.~(\ref{eq:t2}), (\ref{eq:r2}) and (\ref{eq:rt}) corresponding to the output state in Eq.~(\ref{eq:out}) for different values of $\beta_1$, corresponding to the different distances between two coupling points. We set $\beta_1=0$ for the figures of the first row in Fig.~\ref{fig:fig4} as Fig.~\ref{fig:fig4}(a)  to represent the natural atom. We set $\beta_1=\pi/2$ and $\beta_1=2\pi/3$ for the figures of the second and third rows as  Fig.~\ref{fig:fig4}(b) and  Fig.~\ref{fig:fig4}(c) to study the giant atom effect. We find that the two-photon wave function $t_2(x_1,x_2)$ contains only the bound states in this case as shown in  Eq.~(\ref{eq:t2}). Moreover, with the increase of the phase $\beta_1$, the effect of the giant atom become obvious, i.e., the real part of $\Gamma_e$ is decreased, thus the widths  of  $|t_2(x_{1},x_{2})|^2$, $|r_2(x_{1},x_{2})|^2$ and $|r_t(x_{1},x_{2})|^2$ proportional to $1/{\rm Re}(\Gamma_e)$  are gradually increased. That is, the longer distance between two coupling points results in wider distribution of two photons in the real space for the giant atom. Similar to the scattering of the natural atom~\cite{Shen2007PRA}, we find $|r_2(x_{1},x_{2})|^2=0$ for $x_1=x_2$.  This means that two reflected photons cannot reach the same point, i.e., two photons are always anti-bunching. However, the transmitted two photons  are always bunching, because $|t_2(x_{1},x_{2})|^2$ reaches maximum value for $x_1=x_2$.

In Fig.~\ref{fig:fig5} and Fig.~\ref{fig:fig6}, we further show the spatial distributions of $|t_2(x_{1},x_{2})|^2$, $|r_2(x_{1},x_{2})|^2$ and $|r_t(x_{1},x_{2})|^2$ in Eqs.~(\ref{trans:t2}), (\ref{trans:t3}) and (\ref{trans:t4}) for different values of parameters $\delta E_1$ and $\Delta_1$. When the energy difference of two incident photons is equal to the effective decay rate, i.e., $2\Delta_1 =k_1-p_1= -\Gamma(1 + \cos(\beta_1))/2$, and energy detuning satisfies $\delta E_1=2\Delta_L$,  Fig.~\ref{fig:fig5} shows that two transmitted photons can be in anti-bunching when the momenta of two incident photons are different, in contrast to the bunching for two incident photons with the same momentum in Fig.~\ref{fig:fig4}.  This phenomenon is the same as the natural atom case by replacing the effective decay rate $-\Gamma$ for the natural atom with $-\Gamma(1+\cos(\beta_1))/2$ for the giant atom. Besides, the two reflected photons are still anti-bunching no matter what the momenta of two incident photons are. Comparing to the natural atom case for $\beta_1=0$, we observe that the period of oscillation is increased when the phase $\beta_1=k_1\xi_1$ is changed  from $0$ to $\pi/2$ and $3\pi/2$ for the giant atom.
This is because the absolute value of the energy difference $\Delta_1= \Gamma(1 + \cos(\beta_1))/4$ decreases when $\beta_1=k_1\xi_1$ is increased from $0$ to $\pi$. Thus, the contribution of the scattering state, $\cos(\Delta_1x)$, results in an increased period with respect to $x$. When $x$ and $x_c$ are relatively large, the oscillations are influenced only by the scattering state and are unaffected by the bound state because the decay terms $e^{-\Gamma_e|x|/2}\to 0$,  $ e^{-\Gamma_e|x_c|}\to 0$ for large $x$ and $x_c$. Moreover, we find that the wave function $|r_t(x_1,x_2)|^2$,  in which one photon is reflected and another one is transmitted,  becomes asymmetric.  This indicates that the reflected photon leaves the atom earlier than the transmitted photon. This is the same as the case of the two-photon scattering by natural atom.
	
Figure~\ref{fig:fig6} shows the spatial distribution $|t_2(x_{1},x_{2})|^2$, $|r_2(x_{1},x_{2})|^2$ and $|r_t(x_{1},x_{2})|^2$ for $\delta E_1 = 2\Gamma$ and $\Delta_1 = 0$.  When $x$ or $x_c$ is relatively large,  oscillations tend to vanish because the oscillatory term $\cos(\Delta_1 x)$ in the scattering state vanishes and the amplitudes of oscillations from the parts of the bound states tend to zero. However, if  $x$ or $x_c$ is not very large,  compared to the natural atom for $\beta_1=0$,  oscillations in the bound state become more pronounced for the giant atom, e.g., $\beta_1=k_1\xi_1=\pi/2$ and $\beta_1=k_1\xi_1=3\pi/2$. This is because the effective decay rate ${\rm Re}(\Gamma_e)$ given in Eq.~(\ref{eq:Gamma_e}) becomes smaller. Nevertheless, the oscillations eventually disappear for larger value of $x$ or $x_c$  due to the presence of decay. Moreover, for the giant atom, we find that the transmission probability of two-photon increases but the reflection probability of the two-photon decreases  by increasing the phase $\beta_1=k_1\xi_1$ in the region $\beta_1 \in[0,\pi]$ when the total energy $E_{1}$ of two incident photons is constant.
	
To examine the distribution of bound states in momentum space, we express the wave function of the output state in Eq.~(\ref{eq:out})  in momentum space. The transition matrix elements from the incident state $|S^{(R)}_{k_{1},p_{1}}\rangle$ to output state $|S^{(R)}_{k_{2},p_{2}}\rangle$, in which both photons propagate toward the right, is given as
\begin{align}
&\langle S^{(R)}_{k_{2},p_{2}}|S|S^{(R)}_{k_{1},p_{1}}\rangle \nonumber\\
&=\overline{t}_{k_{1}}\overline{t}_{p_{1}} \left[ \delta(k_{1}-k_{2}) \delta(p_{1}-p_{2}) + \delta(k_{1}-p_{2}) \delta(p_{1}-k_{2}) \right] \nonumber\\
&+\frac{1}{4}B \delta(E_{1}-E_{2}),
\end{align}
with the state $|S^{(R)}_{k_{2},p_{2}}\rangle$ given by replacing $k_1$ and $p_1$ with $k_2$ and $p_2$ in Eq.~(\ref{eq:eq61}).
Similarly, the transition matrix elements for both photons propagating toward the left is
\begin{align}
&\langle S^{(L)}_{k_{2},p_{2}}|S|S^{(R)}_{k_{1},p_{1}}\rangle \nonumber\\
&=\overline{r}_{k_{1}}\overline{r}_{p_{1}} \left[ \delta(k_{1}+k_{2}) \delta(p_{1}+p_{2}) + \delta(k_{1}+p_{2}) \delta(p_{1}+k_{2}) \right] \nonumber\\
&+\frac{1}{4}B \delta(E_{1}-E_{2}),
\end{align}
with the state $|S^{(L)}_{k_{2},p_{2}}\rangle$, in which $C^{\dagger}_R(x_1)$ and $C^{\dagger}_R(x_2)$  in Eq.~(\ref{eq:eq61}) are replaced by $C^{\dagger}_L(x_1)$ and $C^{\dagger}_L(x_2)$, and the superscript ``(R)" is replaced by ``(L)". The transition matrix elements for for one photon propagating to the right and another one to the left is
\begin{align}
&\langle k_{2}^{R},p_{2}^{L}|S|S^{(R)}_{k_{1},p_{1}}\rangle
=\overline{t}_{k_{1}}\overline{r}_{p_{1}} \delta(k_{2}-k_{1}) \delta(p_{2}+p_{1})\nonumber\\
&+ \overline{r}_{k_{1}}\overline{t}_{p_{1}} \delta(k_{2}-p_{1}) \delta(p_{2}+k_{1}) +\frac{1}{4}B \delta(E_{1}-E_{2}),
\end{align}
where
\begin{align}
|k_{2}^{R},p_{2}^{L}\rangle=\int dx_1dx_2\frac{1}{2\pi}e^{ik_2x_1+ip_2x_2}C_R^{\dagger}(x_1)C_L^{\dagger}(x_2)|0,-\rangle.
\end{align}
It is clear that the probability amplitude of the bound states scattered in all directions are proportional to $|B|^2$ with $B$ given in Eq.~(\ref{eq:Bk1}).
	
In Fig.~\ref{fig:fig7}, we plot the variations of $|B|^2$ with $\Delta_1=(k_1-p_1)/2$ and $\Delta_2=(k_2-p_2)/2$, which are  the energy difference of two incident photons and the energy difference of two output photons. We set $\beta_1=0$ for figures of the first row in Fig.~\ref{fig:fig7}  to represent the natural atom,  $\beta_1=0.5\pi$ and $\beta_1=0.85\pi$ for the second and third rows in Fig.~\ref{fig:fig7}  to represent the giant atom. By setting the real part of the denominator in Eq.~(\ref{eq:Bk1}) to be zero,  we can have
\begin{align}
\Delta_{1}=\Delta_{2}=\pm\left(\frac{E_1}{2}-\Omega-\frac{\sin(\beta_1)}{4}V^2\right), \label{p1}
\end{align}
which indicates the positions of the peaks in Fig.~\ref{fig:fig7}. The imaginary part of the denominator in Eq.~(\ref{eq:Bk1}), $[1+\cos(\beta_1)]V^2/2$, represents the effective decay rate of the giant atom. When $\beta_1$ is large, the effective decay rate is smaller, resulting in narrower peak widths in Fig.~\ref{fig:fig7}. This implies that the atomic radiation bandwidth is narrower, similar to the reflection coefficient $r_k$ of a single photon.

When the incident energy satisfy the condition $\delta E_1=E_1-2\Omega = \Gamma \sin(\beta_1)/2 = 2\Delta_L$, a peak appears only in the center. This means that two output photons in the bound state tend to be emitted with the same momentum $k_2=p_2$, and the peak value is maximized compared to other energy detuning $\delta E_1$ for each fixed phase. This is consistent with the single photon reflection resonance energy $k = \Omega + \Delta_L$ for the giant atom, differing from $k = \Omega$ for the natural atom.  Compared to the natural atom case for $\beta_1=0$, we find the peak height of the giant atom increases when the phase $\beta_1$ is increased from $0$ to $0.5\pi$ and $0.85\pi$, corresponding to increasing the distance between two coupling points of the giant atom. The analysis above indicates that the giant atom with a smaller decay rate results in a higher peak value for the bound state of the two photons.
	
However, when the energy $\delta E_1\neq 2\Delta_L$ for the giant atom and $\delta E_1\neq 0$ for the natural atom, both the giant atom and the natural atom show four peaks. Specially, in the right column of Fig.~\ref{fig:fig7}, we show the distribution of the bound state for fixed energy detuning $\delta E_1=2\Gamma$. Different from the natural atom case, the peak positions of the giant atom can be tuned by adjusting the phase $\beta_1=k_0\xi_1$ according to Eq.~(\ref{p1}).  The smaller decay rate corresponds to the large distance between two coupling points, thus the widths of these peaks are relatively narrow for the giant atom when distance between two coupling points are large.

\section{Scattering for $N$ coupling points}\label{sec:Nleg}

In this section, we study the two-photon scattering when the giant atom is coupled to the waveguide via $N$ coupling points as schematically shown in Fig.~\ref{fig:fig1}. By using the relation in Eq.~(\ref{eq:dec_eo}), the Hamiltonian in Eq.~(\ref{eq:sysN}) for the giant atom with $N$ coupling points can be rewritten as $H=H_e+H_o$ with $H_o$ given in Eq.~(\ref{eq:eq6}) and $H_e$ given by
\begin{align}\label{eq:eq78}
H_{e} &= \Omega |e\rangle\langle e| -i  v_{g} \int \mathrm{d}x \left[C_{e}^{\dagger}(x) \frac{\partial}{\partial x} C_{e}(x) \right] \\
&\quad +\frac{V}{N}\sum_{i=0}^{N-1} \int \mathrm{d}x   \delta(x - \xi_i) \times \left[ C_e^\dagger(x) \sigma_- + C_e(x) \sigma_+ \right].\nonumber
\end{align}
The coupling strength $V/N$ between the giant atom and the waveguide at $N$ coupling points $0$, $\xi_1$, $\xi_2$, $\ldots$, $\xi_{N-1}$ is the same in the $e$-mode space.  Similar to the case of two coupling points, the interaction takes only place between the giant atom and photons in the $e$-mode space, thus we focus on the derivation of the S-matrix in $e$-mode space.

To obtain the S-matrix for two-photon scattering state, we first calculate the transmission coefficient for a single photon in the $e$-mode space, which can be written as
\begin{equation}\label{eq:tkN}
t_k^N = \frac{2(k - \Omega) - i \Gamma_A^*}{2(k - \Omega) + i \Gamma_A },
\end{equation}
with
\begin{equation}\label{eq:gammaA}
\Gamma_A = \frac{2 V^2}{N^2} \left[ \sum_{j=1}^{N-1} e^{i \beta_j} + \sum_{j=2}^{N-1} \sum_{i=1}^{j-1} e^{-i (\beta_i - \beta_j)} + \frac{N}{2} \right].
\end{equation}
Here, we have used the Markovian approximation with $ k \xi_j = p \xi_j =\Omega \xi_j = \beta_j $,  and $\beta_j$ represents the phase between the coupling point $\xi_0=0$ and $\xi_j$ as shown in Fig.~\ref{fig:fig1}.

Following the procedure in Sec.~\ref{sec:twobethe}, we can obtain the eigenstate of $H_e$ in Eq.~({\ref{eq:eq78}}) in different regions by making a Bethe ansatz (see Appendix~\ref{App:N}). In this case we use $F_l^N(x_1,x_2)$ to denote the probability amplitude of two photons in different regions of the waveguide, and $e_F^N(x)$ to denote the probability amplitude that the giant atom is in the excited state $|e\rangle$ and one photon is in the waveguide, where superscript $``N"$ is used to denote the giant atom with $N$ coupling points. Similar to the  boundary condition in Eq.~({\ref{eq:exboundary}}), we have the relation
\begin{equation}\label{eq:eq81}
e_F^N(0^{-}) e^{i \beta_{N-1}} = e_F^N(\xi_{N-1}^{+}),
\end{equation}
where $\beta_{N-1}$ represents the phase accumulation between the coupling points $\xi_0$ and $\xi_{N-1}$~\cite{Cai2021PRA}. Then, we derive the ratio (see Appendix~\ref{App:N})
\begin{equation}
\frac{A_1}{B_1} = \frac{k - p - i \Gamma_N}{k - p + i \Gamma_N}.
\end{equation}
where $A_1$ and $B_1$ are the coefficients of $F^N_{1,E}(x_2,x_2)$ which is defined in Eq.~(\ref{eq:eqD1}).
Here, the effective decay rate $\Gamma_N$ can be obtained as
\begin{align}\label{eq:gammaN}
\Gamma_N = \frac{V^2}{N^2} \left[ N + \sum_{j=1}^{N-1} 2\cos(\beta_j) +2 \sum_{j=2}^{N-1} \sum_{i=1}^{j-1} \cos (\beta_j - \beta_i)\right].
\end{align}
We can find that the relation $\Gamma_N=\text{Re}(\Gamma_A)$ holds, which is similar to the giant atom with two coupling points $\Gamma_g=\text{Re}(\Gamma_e)$. Then we need to calculate the in-state $|F_{i,E}^N\rangle$  and out-state $|F_{f,E}^N\rangle$ for the solution given by the Bethe ansatz. Using the Lippmann-Schwinger equation, the in-state of the two photons in the $e$-mode space is given by
\begin{equation}
|F_{i,E}^N\rangle = \frac{1}{\sqrt{ 4\Delta^2 + \Gamma_N^2}} \left[ 2\Delta |S_{k,p}\rangle + i \Gamma_N |A_{k,p}\rangle \right],
\end{equation}
where $2\Delta = k - p$ is energy difference of the two incident photons. The out-state can also be obtained as $|F_{f,E}^N\rangle=t_k^Nt_p^N|F_{i,E}^N\rangle$, where $t_k^N$ is given in Eq.~(\ref{eq:tkN}) and $t_p^N$ can be obtained by replacing $k$ with $p$ in Eq.~(\ref{eq:tkN}). Following the same procedure in Sec.~\ref{sec:twobound}, the bound state in the $e$-mode space can be expressed as
\begin{equation}
\left| B_{i,E}^N \right\rangle = \int dx_1 dx_2 B_{i,E}^N(x_1, x_2) \frac{1}{\sqrt{2}} C_e^\dagger(x_1) C_e^\dagger(x_2) | 0, g \rangle,
\end{equation}
with
\begin{equation}
\langle x_{1}, x_{2} | B_{i,E}^N \rangle = B_{i,E}^N(x_{1}, x_{2}) = \sqrt{\frac{\Gamma_N}{4 \pi}} e^{i E x_{c} - \Gamma_{N} |x| / 2}. \label{bo}
\end{equation}
The transmission coefficient for the bound state is
\begin{align}\label{eq:tEN}
t_B^N &= \frac{(E + i \Gamma_N - 2 \Omega) - i \Gamma_{E1}^{*}}{(E - i \Gamma_N - 2 \Omega) + i \Gamma_{E1}}\!\times\! \frac{(E - i \Gamma_N - 2 \Omega) - i \Gamma_{E2}^{*}}{(E + i \Gamma_N - 2 \Omega) + i \Gamma_{E2}},
\end{align}
with
\begin{align}
\Gamma_{E1} &= \frac{V^2}{N^2} \left[ N + 2 \sum_{j=1}^{N} e^{(i \beta_j + \Gamma_N \xi_j / 2)} \right. \nonumber \\
&\quad + 2 \sum_{j=2}^{N-1} \sum_{i=1}^{j-1} e^{-i (\beta_i - \beta_j) + \frac{\Gamma_N}{2} (\xi_j - \xi_i)} \bigg], \nonumber \\
\Gamma_{E2} &= \frac{V^2}{N^2} \left[ N + 2 \sum_{j=1}^{N} e^{(i \beta_j - \Gamma_N \xi_j / 2)} \right. \nonumber \\
&\quad + 2 \sum_{j=2}^{N-1} \sum_{i=1}^{j-1} e^{-i (\beta_i - \beta_j) - \frac{\Gamma_N}{2} (\xi_j - \xi_i)} \bigg],
\end{align}
where $\Gamma_N$ is the effective decay rate defined in Eq.~(\ref{eq:gammaN}).

Under the Markovian approximation, where the delay time between the coupling points is smaller than the atomic decay time, i.e., $\Gamma_N \xi_j \approx 0$ for $j=0,\,\cdots\,,N-1$, we can obtain the two-photon S-matrix $S^{N}_{ee}$ in the $e$-mode space:
\begin{equation}
S^N_{ee} =\sum_{k \leq p} t^N_{k} t^N_{p} |F_{i,E}^N\rangle \langle F_{i,E}^N| + \sum_{E} t^N_{B} |B_{i,E}^N\rangle \langle B_{i,E}^N|.
\end{equation}
where the transmission coefficient $t^N_B$ of the bound state in Eq.~({\ref{eq:tEN}}) under Markovian approximation can be simplified to
\begin{align}
t^N_B &= \frac{E - 2 \Omega - i \Gamma_N - i \Gamma_A^*}{E - 2 \Omega + i \Gamma_N + i \Gamma_A}.
\end{align}
Thus, the total S-matrix of the two-photon scattering for both the $e$-mode and $o$-mode is:
\begin{equation}
S_N = S^N_{ee} + S^N_{oo},\label{eq:SN}
\end{equation}
where $S^N_{oo} $ is the identity matrix in the $o$ space. $S^N_{oo}$ describes the free propagation of photons in the $o$-mode space with
\begin{equation}
S^N_{oo} = \sum |S^{(o)}_{kp}\rangle \langle S^{(o)}_{kp}|,
\end{equation}
which is the same as Eq.~(\ref{eq:60}) for two coupling points.

We now analyze the distribution of the bound state in the momentum space for the output state  when the  incident two-photon state with momentum $k_1$ and $p_1$ is given in Eq.~(\ref{eq:eq61}). Applying the S-matrix $S_N$ in Eq.~(\ref{eq:SN}), we can obtain the terms for the bound state in the momentum space as
\begin{equation}\label{eq:BN}
B_N = \frac{16 i \Gamma_N^2 \Sigma_N}{\pi \left[ 4\Delta_1^2 - (\Sigma_N)^2 \right] \left[ 4\Delta_2^2 - (\Sigma_N)^2 \right]},
\end{equation}
which is very similar to Eq.~(\ref{eq:Bk1}) with
\begin{equation}
\Sigma_N=E_1 - 2 \Omega + i \Gamma_A.
\end{equation}
Here, $2\Delta_1=k_1-p_1$ is the energy difference of two incident photons and $2\Delta_2=k_2-p_2$ is the energy difference of two output photons. $E_1=k_1+p_1$ is the total energy of two incident photons.

To study the effect of the total number $N$ of coupling points, we set the phase between each two coupling points is equal to $\beta$, i.e., $\beta_{j+1}-\beta_{j}=\beta$ for $j=0,\,\cdots\,N-2$, where $\beta_0$ is set as $0$.
Then the imaginary part of $\Gamma_A$ in Eq.~(\ref{eq:gammaA}) corresponding to the Lamb shift can be written as~\cite{Cai2021PRA}
\begin{align}
\Delta^N_L=\frac{\text{Im}(\Gamma_A)}{2}=\frac{V^2}{2N^2}\frac{N\sin(\beta)-\sin(N\beta)}{1-\cos(\beta)}.
\end{align}
The real part of $\Gamma_A$  in Eq.~(\ref{eq:gammaA}) corresponding to the effective decay rate can be written as~\cite{Cai2021PRA}
\begin{align}
\Gamma_N=\text{Re}(\Gamma_A)=\frac{V^2}{N^2}\frac{1-\cos(N\beta)}{1-\cos(\beta)}.
\end{align}
\begin{figure}
\centering
\includegraphics[width=1\linewidth]{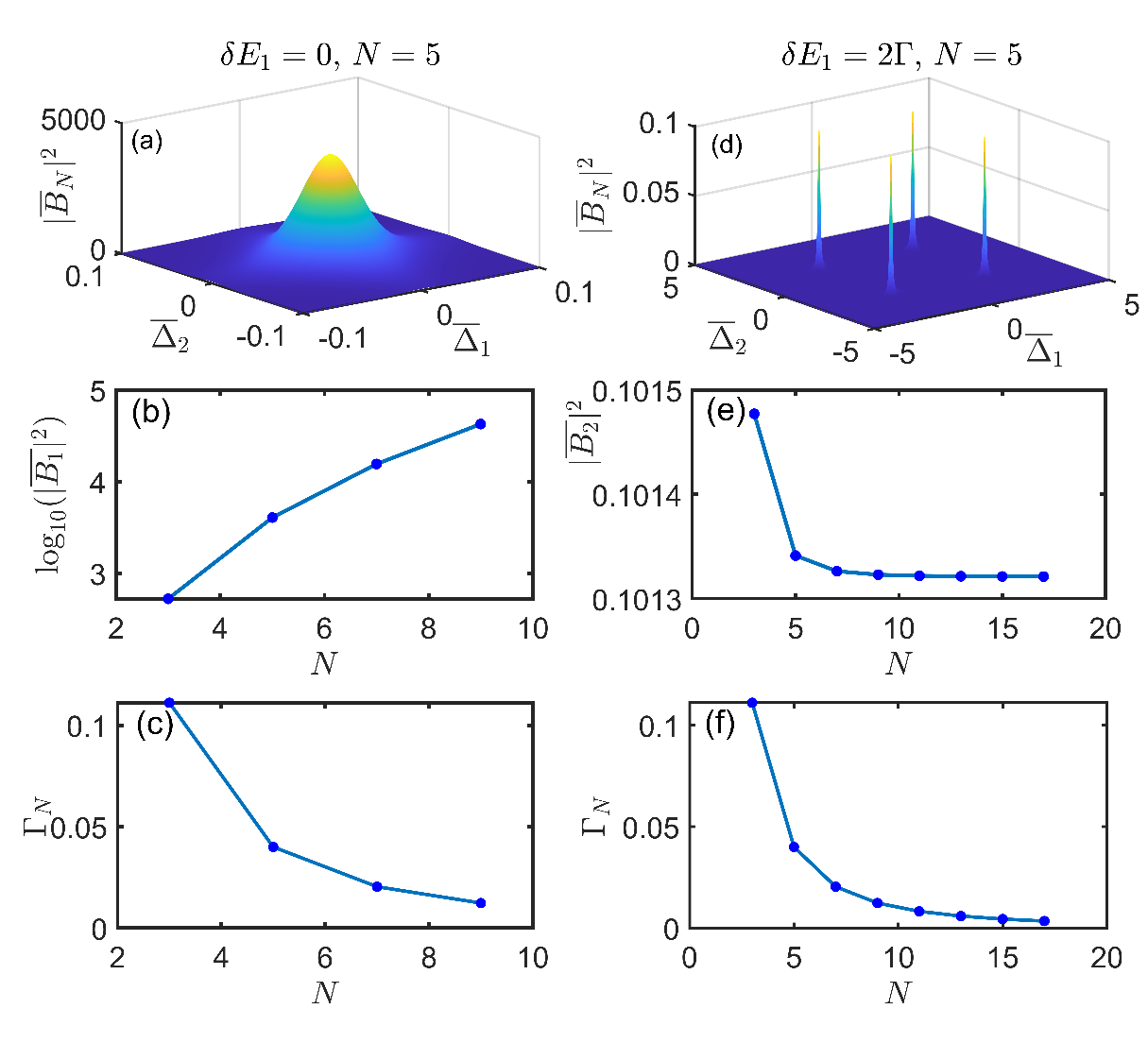}
\caption{(a) and (d) are bound state distribution in momentum space as a function of $\overline{\Delta}_1$ and $\overline{\Delta}_2$, where (a) for energy detuning $\delta E_1=0$ and (b) for $\delta E_1=2\Gamma$. The number of the coupling points of the giant atom is $N=5$. (b) and (e) is the peak value as shown in (a) and (d) for different numbers of the coupling points $N$. (c) and (f) are the effective decay rate for different $N$. In all cases, the phase between each two coupling points is set equally as $\pi$. The other parameters are set as $\overline{\Delta}_{1(2)}= \Delta_{1(2)}/(\Gamma/2)$, $\overline{B}_N = (\Gamma/2)B_N$,$\overline{B_1} = (\Gamma/2)B_1$, $\overline{B_2} = (\Gamma/2)B_2$ and $\delta E_1 = (E_1 - 2\Omega)$, $\Gamma=V^2$.}
\label{fig:fig8}
\end{figure}

In particular,  if we consider the phase $\beta=\pi$, then we obtain the Lamb shift $\Delta^N_L=0$. As discussed for two coupling points case, $\Delta^N_L$ determines the positions of the peaks for bound state distribution in the momentum space as given in Eq.~(\ref{eq:BN}). So the peak must locate at $\Delta_1=\pm {\rm Re} \sum_N/2 $ and $\Delta_2=\pm {\rm Re} \sum_N/2$. Moreover, if we assume that the energy of each photon satisfies the condition $k_1=p_1=\Omega$, which results in $\delta E_1=E_1-2\Omega=0$. Then, there is a maximum distribution around the peak with the height $B_1$ written as
\begin{align}\label{eq:B1}
B_1=-\frac{16 }{\pi\Gamma_N}.
\end{align}
which is obtained from Eq.~(\ref{eq:BN}) by setting $\Delta_1=\Delta_2=0$ and $E_1=2\Omega$.

In Fig.~\ref{fig:fig8}(a), we plot the bound state distribution in the momentum space with the phase $\beta=\pi$. We find the two  photons in the output state tend to have the same momentum at the peak, which locates around $\Delta_1= \Delta_2=0$. Figure~\ref{fig:fig8}(b) shows the peak value $|B_1|^2$ in Eq.~(\ref{eq:B1}) at $\Delta_1=\Delta_2=0$ increases in a $N^4$ manner with the total number $N$ of coupling points. Moreover, the total number $N$ can only be odd number, because the giant atom is decoupled from the waveguide when $N$ is even for the phase $\beta=\pi$. Figure~\ref{fig:fig8}(c) demonstrates that the effective decay $\Gamma_N$ decays in $N^2$. That is, the width of the peak in Fig.~\ref{fig:fig8}(a) narrows with $N^2$.

We now study  the case that $\delta E_1=E_1-2\Omega\neq0$,  e.g., $\delta E_1=2\Gamma$. Similar to the giant atom with two coupling points, there are four peaks determined by:
\begin{align}
\Delta_1=\Delta_2=\pm\frac{1}{2}\left(E_1-2\Omega-2\Delta_L\right).
\end{align}
Without loss of generality, we chose a peak located at $\Delta_1=\Delta_2=(E_1-2\Omega-2\Delta_L)/2$ and substitute them into Eq.~(\ref{eq:BN}).
Then, the height of the peak can be obtained as
\begin{align}\label{eq:B2}
B_2=\frac{-16 \Sigma_N}{\pi \left[2\Delta_1+\Sigma_N\right] \left[2\Delta_2+\Sigma_N \right]}.
\end{align}
In the Fig.~\ref{fig:fig8}(d), we plot the bound state distribution for $\delta E_1=2\Gamma$ with the phase $\beta=\pi$. Different from $\delta E_1=0$, this distribution shows four peaks with the same height. This indicates that the proportion of bound states in the output state decreases when the single-photon resonance condition is no longer met. In Fig.~\ref{fig:fig8}(e),  the height for one of four peaks given in Eq.~(\ref{eq:B2}) is plotted as the function of the number $N$ of the coupling points. We find that the height of the peak decreases with $N$ and gradually tends to a constant when $N$ becomes large. Similar to the case $\delta E_1=0$, the total number $N$ can only be odd because the effective decay rate $\Gamma_N$ is the same for  $\delta E_1=0$ and  $\delta E_1\neq0$. The widths of four peaks determined by the decay rate $\Gamma_N$ are also the function of $N^2$ as shown in Fig.~\ref{fig:fig8}(f). Combining Fig.~\ref{fig:fig8}(e) and Fig.~\ref{fig:fig8}(f), we find the proportion of bound states in the output state also decreases with the increase of $N$.

\section{Conclusions and discussions}\label{sec:conclusion}

In conclusion, we have studied the two-photon scattering in a waveguide which is coupled to a giant atom via two or $N$ spatial points.  We first study the case that the giant atom is coupled to the waveguide via two coupling points. We use the real-space Bethe ansatz method to determine the interacting eigenstates of the system under the Markovian approximation,  and then use these eigenstates and eigenvalues to calculate the in-state and out-state via Lippmann-Schwinger equation.  Since the in-state given by Bethe ansatz solution is not complete, we obtain the other in-state as bound state via the Bethe ansatz solutions. Thus, we obtain all the eigenstate and eigenvalues of S-matrix  and construct the S-matrix for the two-photon scattering by a giant atom, which is coupled to the waveguide via two coupling points.  These eigenstates and eigenvalues have never been given by others.

We study the properties of output states for given incident two-photon states via the two-photon S-matrix. As an example, we choose two-photon incident state in $R$-mode space described in Eq.~(\ref{eq:eq61}). We find that the oscillation period of the scattering states can be increased or decreased by varying the distance between two coupling points. The decay rate of the two-photon bound states can also be enhanced or reduced by varying the distance between two coupling points. By using these variations, we can change the bound state distribution in the real space for two photon transmitted wave function under single photon resonance condition. This cannot be achieved in the two-photon scattering by a natural atom. Moreover, for the two-photon bound states, we find that the momentums of  two emitted photons tend to be identical when each of two incident single photons satisfies the single-photon resonant scattering condition. Thus, there is  one peak in the momentum space. Compared to the natural atom, the peak height is enhanced largely which indicates the photon correlation can be enhanced by the giant atom. For non-resonant scattering, the bound states have four peaks in the momentum space, which is similar to that of natural atoms. But the width of the peak can be reduced by increasing the distance between coupling points, this is because the effective decay rate is decreased with the increase of the distance between coupling points.  Compared to the work in~\cite{Gu2023PRA} that only considers two incident photons with equal energy, we study all possibilities of energy difference for two incident photons when bound state distribution is calculated for a given total energy. Moreover, we calculate the momentum distribution of bound state for the output states which is not derived in~\cite{Gu2023PRA}. We find that the transmitted wave function is still bunching when two incident photons satisfy the single photon resonance condition. We also find that both transmitted and reflected photons can be in anti-bunching in some parameter regimes for the non-resonant scattering.
	
We extend our study to the case that the giant atom is coupled to the waveguide via $N$ coupling points. We obtain the eigenstates and eigenvalues of S-matrix and construct the S-matrix under the Markovian approximation. We mainly study the case that the phases between any two nearest neighbor coupling points are $\pi$. We find the height value of the peak increases with $N^4$ when each of two incident single photon satisfies the resonant scattering condition. Thus, the photon correlation can be enhanced by increasing the coupling points number. For non-resonant scattering case, the momentum distribution of bound states has four peaks with the same height which remains a constant when the number $N$ of the coupling points is large. In both cases, the width of the peak decrease with $N^2$ when the distance between two nearest neighbor coupling points is fixed.

 Finally, we discuss the experimental feasibility for our study. The giant atom formed by the superconducting qubit circuit has recently been coupled to the transmission line waveguide via two~\cite{Kannan2020N} or six coupling points~\cite{Vadiraj2021PRA}. In these experiments, superconducting qubit and waveguide is in weak coupling regime, thus the rotating-wave approximation used for our Hamiltonian is valid~\cite{Vadiraj2021PRA}. Moreover, the smallest relaxation time of the qubit induced by the waveguide is $10^{-1}\mu s$~\cite{Kannan2020N}, which is significantly larger than the traveling time of the light between coupling points $10^{-5}\mu s$~\cite{Vadiraj2021PRA}. Thus, the Markovian approximation is valid. That is,  our study  is experimentally feasible for current technology of superconducting quantum circuits.

\appendix
\section{Probability amplitude for Bethe ansatz method}\label{App:scattering}
In this appendix, we provide a detailed derivation of the wave function coefficients $ g(x_1,x_2) $ and $ e(x) $ solved by Bethe ansatz method.
Using Eq.~(\ref{eq:boundary_gx}), we can determine the boundary conditions for $ g(x_1,x_2) $ which applies in both case for the solution given by Bethe ansatz and solution given by bound state. The boundary condition between regions $\mathcal{F}_1$ and $\mathcal{F}_2$ illustrated in Fig.~\ref{fig:fig3} is given by
\begin{equation}\label{eq:A1}
-i\left[g(x_1,0^+) - g(x_1,0^-)\right] + \frac{V}{2\sqrt{2}} e(x_1) = 0,
\end{equation}
between regions $\mathcal{F}_2$ and $\mathcal{F}_3$ is
\begin{equation}
-i\left[g(x_1,\xi_1^+) - g(x_1,\xi_1^-)\right] + \frac{V}{2\sqrt{2}} e(x_1) = 0, \label{eq:BC}
\end{equation}
between regions $\mathcal{F}_3$ and $\mathcal{F}_5$ is
\begin{equation}
-i\left[g(0^+,x_2) - g(0^-,x_2)\right] + \frac{V}{2\sqrt{2}} e(x_2) = 0,
\end{equation}
and between regions $\mathcal{F}_5$ and $\mathcal{F}_6$ is
\begin{equation}\label{eq:A4}
-i\left[g(\xi_1^+,x_2) - g(\xi_1^-,x_2)\right] + \frac{V}{2\sqrt{2}} e(x_2) = 0.
\end{equation}
For the Bethe ansatz case, we substitute $g(x_1,x_2)=F_{l,E}(x_1,x_2)$  for $l=1,\,\cdots ,\,6$ in Eq.~(\ref{eq:eq30}) and $e(x)=e_F(x)$ into above boundary equations. Combining with Eq.~(\ref{eq:boundary_ex}), we obtain the following corresponding eight equations for coefficients $A_l$ and $B_l$ as:
\begin{align}
(\Omega - p)(A_{2} - A_{1}) &= i\frac{V^{2}}{8} \left[(A_{1} + A_{2}) + (A_{2} + A_{3}) e^{ip\xi_1}\right], \nonumber\\
(\Omega - p)(A_{3} - A_{2}) &= i\frac{V^{2}}{8} \left[(A_{1} + A_{2}) e^{-ip\xi_1} + (A_{2} + A_{3})\right], \nonumber\\
(\Omega - k)(A_{5} - A_{3}) &= i\frac{V^{2}}{8} \left[(A_{3} + A_{5}) + (A_{5} + A_{6}) e^{ik\xi_1}\right], \nonumber\\
(\Omega - k)(A_{6} - A_{5}) &= i\frac{V^{2}}{8} \left[(A_{3} + A_{5}) e^{-ik\xi_1} + (A_{5} + A_{6})\right], \label{eq:A}
\end{align}
and
\begin{align}
(\Omega - k)(B_{2} - B_{1}) &= i\frac{V^{2}}{8} \left[(B_{1} + B_{2}) + (B_{2} + B_{3}) e^{ik\xi_1}\right], \nonumber\\
(\Omega - k)(B_{3} - B_{2}) &= i\frac{V^{2}}{8} \left[(B_{1} + B_{2}) e^{-ik\xi_1} + (B_{2} + B_{3})\right], \nonumber\\
(\Omega - p)(B_{5} - B_{3}) &= i\frac{V^{2}}{8} \left[(B_{3} + B_{5}) + (B_{5} + B_{6}) e^{ip\xi_1}\right], \nonumber\\
(\Omega - p)(B_{6} - B_{5}) &= i\frac{V^{2}}{8} \left[(B_{3} + B_{5}) e^{-ip\xi_1} + (B_{5} + B_{6})\right].
\end{align}
Solving these eight equations, we obtain the ratio:
\begin{align}
\frac{A_{2}}{A_{1}} &= \frac{4(p - \Omega)}{4(p - \Omega) + iV^{2}(1 + e^{ip\xi_1})}, \nonumber\\
\frac{A_{3}}{A_{1}} &= \frac{4(p - \Omega) - iV^{2}(1 + e^{-ip\xi_1})}{4(p - \Omega) + iV^{2}(1 + e^{ip\xi_1})}, \nonumber\\
\frac{A_{5}}{A_{3}} &= \frac{4(k - \Omega)}{4(k - \Omega) + iV^{2}(1 + e^{ik\xi_1})},
\end{align}
and
\begin{align}
\frac{B_{2}}{B_{1}} &= \frac{4(k - \Omega)}{4(k - \Omega) + iV^{2}(1 + e^{ik\xi_1})}, \nonumber\\
\frac{B_{3}}{B_{1}} &= \frac{4(k - \Omega) - iV^{2}(1 + e^{-ik\xi_1})}{4(k - \Omega) + iV^{2}(1 + e^{ik\xi_1})}, \nonumber\\
\frac{B_{5}}{B_{3}} &= \frac{4(p - \Omega)}{4(p - \Omega) + iV^{2}(1 + e^{ip\xi_1})}.
\end{align}
We also have the relation $A_6/A_1 = B_6/B_1 = t_k t_p$, where $t_k$ and $t_p$ refer to the transmission coefficient of a single photon, as given in Eq.~(\ref{eq:tk1}).

As mentioned in the main text, while we have known the ratio coefficients between different regions, determining the ratio between $ A_1 $ and $ B_1 $ is key to solving the eigenstates. Therefore, we now derive the explicit form of the probability amplitude $ e_F(x) $ that there is one photon in the waveguide and the giant atom is in the excited state $|e\rangle$.
	
For $x<0$, according to the Eq.~(\ref{eq:boundary_ex}), the boundary condition is given by
\begin{align}
\left(i\frac{\partial}{\partial x_1}-\Omega+E\right)e_F(x)&=V^{\prime}\sum_{m=0}^1\bigg[g(x,\xi_m^{-})+g(x,\xi_m^{+})\bigg],\label{eq:ex}
\end{align}
where we set $V^{\prime}=V/(2\sqrt{2})$ and $\xi_0=0$. By substituting Eq.~(\ref{eq:A1}) and Eq.~(\ref{eq:BC}) into the above Eq.~(\ref{eq:ex}), we obtain
\begin{align}
i\frac{\partial}{\partial x}e_F(x)=\left[\Omega-E-\frac{iV^{2}}{4}\right]e_F(x)+2V^{\prime}\sum_{m=0}^1g(x,\xi_m^{-}).
\end{align}
We integrate the above equation with respect to $x$ and obtain:
\begin{align}
&e_F(x) =ie^{i\left(E-\Omega+\frac{iV^{2}}{4}\right)x}\int_{-\infty}^{x}\mathrm{d}x^{\prime}e^{-i\left(E-\Omega+\frac{iV^{2}}{4}\right)x^{\prime}} \nonumber\\
&\times\left(\frac{-V}{\sqrt{2}}\right)\left[\left(A_1+A_2e^{ip\xi_1}\right)e^{ikx^{\prime}}+\left(B_1+B_2e^{ik\xi_1}\right)e^{ipx^{\prime}}\right] \nonumber\\
&=\frac{V\left(A_1+A_2e^{ip\xi_1}\right)}{\sqrt{2}\left(p-\Omega+\frac{iV^2}{4}\right)}e^{ikx}+\frac{V\left(B_1+B_2e^{ik\xi_1}\right)}{\sqrt{2}\left(k-\Omega+\frac{iV^2}{4}\right)}e^{ipx}.\label{ex0}
\end{align}
Similarly, for $0<x<\xi_1$, we have
\begin{align}
i\frac{\partial}{\partial x}e_F(x)=\left[\Omega-E-\frac{iV^{2}}{4}\right]e_F(x)+\frac{V\left[g(0^{-},x)+g(x,\xi_1^{-})\right]}{\sqrt{2}}.
\end{align}
Thus, the probability amplitude $e_F(x)$ can be expressed as
\begin{align}
&e_F(x) =ie^{i\left(E-\Omega+\frac{iV^{2}}{4}\right)x}\int_{-\infty}^{x}\mathrm{d}x^{\prime}e^{-i\left(E-\Omega+\frac{iV^{2}}{4}\right)x^{\prime}} \nonumber\\	&\times\left(\frac{-V}{\sqrt{2}}\right)\left[\left(B_2+A_4e^{ip\xi_1}\right)e^{ikx^{\prime}}+\left(A_2+B_4e^{ik\xi_1}\right)e^{ipx^{\prime}}\right] \nonumber\\
&=\frac{V\left(B_2+A_4e^{ip\xi_1}\right)}{\sqrt{2}\left(p-\Omega+\frac{iV^2}{4}\right)}e^{ikx}+\frac{V\left(A_2+B_4e^{ik\xi_1}\right)}{\sqrt{2}\left(k-\Omega+\frac{iV^2}{4}\right)}e^{ipx}.
\end{align}
Finally, for $x>\xi_1$, we have
\begin{align}
i\frac{\partial}{\partial x}e_F(x)=\left[\Omega-E-\frac{iV^{2}}{4}\right]e_F(x)+2V^{\prime}\sum_{m=0}^1g(\xi_m^{-},x).
\end{align}
Thus:
\begin{align}
&e_F(x) =ie^{i\left(E-\Omega+\frac{iV^{2}}{4}\right)x}\int_{-\infty}^{x}\mathrm{d}x^{\prime}e^{-i\left(E-\Omega+\frac{iV^{2}}{4}\right)x^{\prime}} \nonumber\\
&\times\left(\frac{-V}{\sqrt{2}}\right)\left[\left(A_3+A_5e^{ip\xi_1}\right)e^{ikx^{\prime}}+\left(B_3+B_5e^{ik\xi_1}\right)e^{ipx^{\prime}}\right] \nonumber\\
&=\frac{V\left(A_3+A_5e^{ip\xi_1}\right)}{\sqrt{2}\left(p-\Omega+\frac{iV^2}{4}\right)}e^{ikx}+\frac{V\left(B_3+B_5e^{ik\xi_1}\right)}{\sqrt{2}\left(k-\Omega+\frac{iV^2}{4}\right)}e^{ipx}.\label{exx0}
\end{align}

\section{Derivation of the bound state and probability amplitude for eigenstate in Eq.~(\ref{eq:bound})}\label{App:bound}
In this appendix, we provide a detailed derivation of the bound state wave function and another eigenstate of $H_e$ which corresponds to probability amplitude $ B_{l,E}(x_1,x_2) $ and $ e_B(x) $. To derive the bound state, we subtract the scattering state part from the entire space to obtain the complementary space. So we calculate the expression:
	
\begin{equation}
|\delta_{k_1,p_1}\rangle \equiv |S^{(e)}_{k_1,p_1}\rangle - \sum_{k\leq p} \, \langle F_{i,E}|S^{(e)}_{k_1,p_1}\rangle |F_{i,E}\rangle,
\end{equation}
where $|S^{(e)}_{k_1,p_1}\rangle$ can be obtained by setting $k=k_1$ and $p=p_1$ in Eq.~(\ref{eq:Se}). Here, $|F_{i,E}\rangle$ is defined in Eq.~(\ref{sca}) with $E=k+p$. Since our calculations are performed in the $e$-mode space, we will omit the superscript $``(e)"$ below. First we calculate the overlap between scattering state $|F_{i,E}\rangle$ and the entire space $|S_{k_{1},p_{1}}\rangle$,
\begin{align}
&\langle F_{i,E}|S_{k_{1},p_{1}}\rangle\nonumber\\
&= \frac{1}{\sqrt{4\Delta^{2} + \Gamma_g^{2}}} \left( 2\Delta \langle S_{k,p}|S_{k_{1},p_{1}}\rangle- i\Gamma_g \langle A_{k,p}|S_{k_{1},p_{1}}\rangle \right)\nonumber\\
&= \frac{1}{\sqrt{4\Delta^{2} + \Gamma_g^{2}}} \biggl( 2\Delta \left[ \delta(\Delta - \Delta_{1}) + \delta(\Delta + \Delta_{1}) \right]\nonumber\\
&\quad - 2\Delta \frac{\Gamma_g}{\pi} \mathcal{P} \frac{1}{\Delta^{2} - \Delta_{1}^{2}} \biggr) \delta(E - E_{1}),
\end{align}
	
Here, $2\Delta=k-p$ is the energy difference of the two photons. The overlap $\langle S_{k,p}|S_{k_{1},p_{1}}\rangle$ and $\langle A_{k,p}|S_{k_{1},p_{1}}\rangle$ are given in Ref.~\cite{Shen2007PRA}.
We then project $| \delta_{k_{1},p_{1}}\rangle$ onto the $|S_{k_{2},p_{2}}\rangle$ space and calculate
	
\begin{align}
\langle &S_{k_{2},p_{2}}| \delta_{k_{1},p_{1}}\rangle \nonumber\\
&= \langle S_{k_{2},p_{2}}|S_{k_{1},p_{1}}\rangle - \sum_{k\leq p} \langle F_{i,E}|S_{k_{1},p_{1}}\rangle \langle S_{k_{2},p_{2}}|F_{i,E}\rangle.
\end{align}
	
Since the integral is symmetric with respect to $ k = 0 $ and $ p = 0 $, we have:
	
\begin{align}\label{eq:B4}
\sum_{k,p,k \leq p} \langle& S_{k_{2},p_{2}}|F_{i,E}\rangle \langle F_{i,E}|S_{k_{1},p_{1}}\rangle \nonumber\\
&= \frac{1}{2} \sum_{k,p} \langle S_{k_{2},p_{2}}|F_{i,E}\rangle \langle F_{i,E}|S_{k_{1},p_{1}}\rangle.
\end{align}
Then we calculate the overlap $\langle S_{k_{2},p_{2}}|F_{i,E}\rangle \langle F_{i,E}|S_{k_{1},p_{1}}\rangle $ which is a crucial step to solving the bound state. We find that an analytical expression for the bound state can only be obtained under the Markov approximation.

\begin{align}
&\langle S_{k_{2},p_{2}}|F_{i,E}\rangle \langle F_{i,E}|S_{k_{1},p_{1}}\rangle \nonumber\\
&= \frac{4\Delta^2}{4\Delta^{2} + \Gamma_g^{2}} \delta(E - E_{1}) \delta(E - E_{2}) \times \biggl\{\Delta_1^{\prime}\Delta_2^{\prime}\nonumber\\
&+\left( -\frac{\Gamma_g}{\pi} \right) \Delta_1^{\prime} \mathcal{P} \frac{1}{\Delta^{2} - \Delta_{2}^{2}} + \left( -\frac{\Gamma_g}{\pi} \right) \Delta_2^{\prime} \mathcal{P} \frac{1}{\Delta^{2} - \Delta_{1}^{2}} \nonumber\\
&+ \left( \frac{\Gamma_g}{\pi} \right)^{2} \mathcal{P} \frac{1}{\Delta^{2} - \Delta_{1}^{2}} \mathcal{P} \frac{1}{\Delta^{2} - \Delta_{2}^{2}} \biggr\}
\end{align}
where $\mathcal{P}$ represents the Cauchy principal value integral. Here, we have set $\Delta_{1(2)}^{\prime}=\delta\left(\Delta-\Delta_{1(2)}\right)+\delta\left(\Delta+\Delta_{1(2)}\right)$. Further simplification yields:
		
\begin{align}
&\langle S_{k_{2},p_{2}}|F_{i,E}\rangle \langle F_{i,E}|S_{k_{1},p_{1}}\rangle \nonumber\\
&= \delta(E - E_{1}) \delta(E - E_{2}) \times \biggl\{ \Delta_1^{\prime}\Delta_2^{\prime}  \nonumber\\
&+ \left( -\frac{\Gamma_g}{\pi} \right) \frac{4\Delta_1^2}{4\Delta^{2} + \Gamma_g^{2}} \mathcal{P} \frac{1}{\Delta_{1}^{2} - \Delta_{2}^{2}} \Delta_1^{\prime} \nonumber\\
&+ \left( -\frac{\Gamma_g}{\pi} \right) \frac{4\Delta_2^2}{4\Delta^{2} + \Gamma_g^{2}} \mathcal{P} \frac{1}{\Delta_{2}^{2} - \Delta_{1}^{2}} \Delta_2^{\prime} \nonumber\\
&+ \left( \frac{\Gamma_g}{\pi} \right)^{2} \frac{4\Delta^2}{4\Delta^{2} + \Gamma_g^{2}} \mathcal{P} \frac{1}{\Delta_{1}^{2} - \Delta_{2}^{2}}  \mathcal{P} \frac{\Delta_1^2-\Delta_2^2}{(\Delta^{2} - \Delta_{1}^{2})(\Delta^{2} - \Delta_{2}^{2})}\biggr\}.
\end{align}
Next, we perform the integration over $ k $ and $ p $ from $-\infty$ to $\infty$, then the overlap in Eq.~(\ref{eq:B4}) can be obtained as
\begin{align}
&\frac{1}{2} \sum_{k,p} \langle S_{k_{2},p_{2}}|F_{i,E}\rangle \langle F_{i,E}|S_{k_{1},p_{1}}\rangle\nonumber\\
&=  \int_{-\infty}^{+\infty} dE \int_{-\infty}^{+\infty} d\Delta \, \delta(E - E_{1}) \delta(E - E_{2}) \times \biggl\{ \Delta_1^{\prime}\Delta_2^{\prime} \nonumber\\
&+ \left( -\frac{\Gamma_g}{\pi} \right) \frac{4\Delta_1^2}{4\Delta^{2} + \Gamma_g^{2}} \mathcal{P} \frac{1}{\Delta_{1}^{2} - \Delta_{2}^{2}} \Delta_1^{\prime} \nonumber\\
&+ \left( -\frac{\Gamma_g}{\pi} \right) \frac{4\Delta_2^2}{4\Delta^{2} + \Gamma_g^{2}} \mathcal{P} \frac{1}{\Delta_{2}^{2} - \Delta_{1}^{2}} \Delta_2^{\prime} \nonumber\\
&+ \left( \frac{\Gamma_g}{\pi} \right)^{2} \frac{4\Delta^2}{4\Delta^{2} + \Gamma_g^{2}} \mathcal{P} \frac{1}{\Delta_{1}^{2} - \Delta_{2}^{2}} \mathcal{P} \frac{\Delta_1^2-\Delta_2^2}{(\Delta^{2} - \Delta_{1}^{2})(\Delta^{2} - \Delta_{2}^{2})} \biggr\}.
\end{align}

Fortunately, under the Markov approximation, the last term can be calculated analytically using a contour integral. Then, following the same process as in reference~\cite{Shen2007PRA}, under the Markovian approximation, the bound state can be expressed as:
\begin{equation}
\left|B_{i,E}\right\rangle=\int dx_1 dx_2 \, B_{i,E}(x_1, x_2) \frac{1}{\sqrt{2}} C_e^\dagger(x_1) C_e^\dagger(x_2) |0, g \rangle,	
\end{equation}
where
\begin{equation}
\langle x_1, x_2 | B_{i,E} \rangle = B_{i,E}(x_1, x_2) = \sqrt{\frac{\Gamma_g}{4\pi}} e^{iEx_c - \Gamma_g |x|/2},
\end{equation}
	
Having already determined the bound state, we can now calculate the transmission coefficient for bound state by assuming a wave function as described in Eq.~(\ref{eq:bound}) and applying the boundary conditions in Eq.~(\ref{eq:A1}) to (\ref{eq:A4}). For the bound state case, we substitute $g(x_1,x_2)=B_{l,E}(x_1,x_2)$  for $l=1,\,\cdots ,\,6$ in Eq.~(\ref{eq:bound}) and $e(x)=e_B(x)$ into boundary equations and combine with Eq.~(\ref{eq:boundary_ex}). The resulting equations, which are analogous to Eq.~(\ref{eq:A}), are derived as follows:
\begin{align}
&8(t_2 - t_1)\left(2i\Omega + \Gamma_g - E\right) + V^2(t_2 + t_1) \nonumber\\
&+ V^2(t_2 + t_3) e^{(iE - \Gamma_g)\xi_1 / 2} = 0 \\
&8(t_3 - t_2)(2i\Omega + \Gamma_g - E) e^{(iE - \Gamma_g)\xi_1 / 2} \nonumber\\
&+ V^2(t_2 + t_1) + V^2(t_2 + t_3) e^{(iE - \Gamma_g)\xi_1 / 2} = 0 \\
&8(t_5 - t_3)(2i\Omega -\Gamma_g - E) + V^2(t_3 + t_5) \nonumber\\
&+ V^2(t_5 + t_6) e^{(iE - \Gamma_g)\xi_1 / 2} = 0 \\
&8(t_6 - t_5)(2i\Omega - \Gamma_g - E) e^{(iE + \Gamma_g)\xi_1 / 2} \nonumber\\
&+ V^2(t_3 + t_5) + V^2(t_5 + t_6) e^{(iE - \Gamma_g)\xi_1 / 2} = 0
\end{align}
Solving the equations above, we have the transmission coefficient between region $\mathcal{F}_1$ and $\mathcal{F}_2$ illustrated in Fig.~\ref{fig:fig3} as
\begin{align}
t_{2} &= \frac{2(E + i\Gamma_{g} - 2\Omega)}{2E + 2i\Gamma_{g} - 4\Omega + iV^2 \left(1 + e^{\frac{iE - \Gamma_{g}}{2} \xi_1} \right)},
\end{align}
and the transmission coefficient between region $\mathcal{F}_1$ and $\mathcal{F}_3$ as
\begin{align}
t_{3} &= \frac{2E + 2i\Gamma_{g} - 4\Omega - iV^2 \left(1 + e^{\frac{-iE + \Gamma_{g}}{2} \xi_1} \right)}{2E + 2i\Gamma_{g} - 4\Omega + iV^2 \left(1 + e^{\frac{iE - \Gamma_{g}}{2} \xi_1} \right)} ,
\end{align}
and the transmission coefficient between region $\mathcal{F}_1$ and $\mathcal{F}_5$ as
\begin{align}
t_{5} &= \frac{2E + 2i\Gamma_{g} - 4\Omega - iV^2 \left(1 + e^{\frac{-iE + \Gamma_{g}}{2} \xi_1} \right)}{2E - 2i\Gamma_{g} - 4\Omega + iV^2 \left(1 + e^{\frac{iE - \Gamma_{g}}{2} \xi_1} \right) } \nonumber \\
&\times \frac{2(E - i\Gamma_{g} - 2\Omega)}{2E + 2i\Gamma_{g} - 4\Omega + iV^2 \left(1 + e^{\frac{iE + \Gamma_{g}}{2} \xi_1} \right)} .
\end{align}
Finally we obtain the transmission coefficient for both photons passing through the two coupling points of the giant atom.
\begin{align}\label{eq:t6}
t_{6} &= \frac{2E - 4\Omega + 2i\Gamma_{g} - iV^2 \left(1 + e^{\frac{-iE + \Gamma_{g}}{2} \xi_1} \right)}{2E - 4\Omega - 2i\Gamma_{g} + iV^2 \left(1 + e^{\frac{iE + \Gamma_{g}}{2} \xi_1} \right)} \nonumber \\
&\times \frac{2E - 4\Omega - 2i\Gamma_{g} - iV^2 \left(1 + e^{\frac{-iE - \Gamma_{g}}{2} \xi_1} \right)}{2E - 4\Omega + 2i\Gamma_{g} + iV^2\left(1 + e^{\frac{iE - \Gamma_{g}}{2} \xi_1} \right)}, \nonumber \\
\end{align}
where we have set $t_1=1$ without loss of generality.
Next, we will use $ B_{l,E}(x_1,x_2) $ to determine $ e_E(x) $ for each regions similar to the Bethe ansatz method.
When $x < 0$, we obtain:
\begin{align}
e_B(x) &= i e^{i\left(E - \Omega + \frac{iV^{2}}{4}\right)x}
\int_{-\infty}^{x} \mathrm{d}x^{\prime}
e^{-i\left(E - \Omega + \frac{iV^{2}}{4}\right)x^{\prime}} \nonumber \\
&\quad \times \left(-\frac{V}{\sqrt{2}}\right) \left[
e^{\left(iE + \Gamma_g\right)\frac{x}{2}}
+ t_2 e^{\left(iE + \Gamma_g\right)\frac{\xi_1}{2}} e^{\left(iE - \Gamma_g\right)\frac{\xi_1}{2}} \right] \nonumber \\
&= \frac{V}{\sqrt{2}} \frac{e^{\left(iE + \Gamma_g\right)\frac{x}{2}}}{\frac{E}{2} - \Omega + \frac{i\Gamma_g}{2} + \frac{iV^2}{4}}
\left(1 + t_2 e^{\left(iE - \Gamma_g\right)\frac{\xi_1}{2}} \right) \label{eq:eEx}
\end{align}
When $0 < x < \xi_1$, we obtain:
\begin{align}
&e_B(x) = i e^{i\left(E - \Omega + \frac{iV^{2}}{4}\right)x}
\int_{-\infty}^{x} \mathrm{d}x^{\prime}
e^{-i\left(E - \Omega + \frac{iV^{2}}{4}\right)x^{\prime}} \nonumber \\
&\quad \times \left(-\frac{V}{\sqrt{2}}\right) \left[
t_2 e^{\left(iE - \Gamma_g\right)\frac{x}{2}}
+ t_4 e^{\left(iE + \Gamma_g\right)\frac{x}{2}} e^{\left(iE - \Gamma_g\right)\frac{\xi_1}{2}} \right] \nonumber \\
&= \frac{V}{\sqrt{2}} \left(
\frac{t_2 e^{\left(iE - \Gamma_g\right)\frac{x}{2}}}{\frac{E}{2} - \Omega - \frac{i\Gamma_g}{2} + \frac{iV^2}{4}}
+ \frac{t_4 e^{\left(iE - \Gamma_g\right)\frac{\xi_1}{2}} e^{\left(iE + \Gamma_g\right)\frac{x}{2}}}{\frac{E}{2} - \Omega + \frac{i\Gamma_g}{2} + \frac{iV^2}{4}}
\right)
\end{align}
When $x>\xi_1$, we obtain:
\begin{align}
e_B(x)& =ie^{i\left(E-\Omega+\frac{iV^{2}}{4}\right)x}\int_{-\infty}^{x}\mathrm{d}x^{\prime}e^{-i\left(E-\Omega+\frac{iV^{2}}{4}\right)x^{\prime}} \nonumber\\
		&\times\left(-\frac{V}{\sqrt{2}}\right)\left[t_3e^{(iE-\Gamma_g)\frac{x}{2}}+t_5e^{(iE-\Gamma_g)\frac{x}{2}}e^{(iE+\Gamma_g)\frac{\xi_1}{2}}\right] \nonumber\\
		&=\frac{V}{\sqrt{2}}\frac{e^{\left(iE-\Gamma_{g}\right)\frac{x}{2}}}{\frac{E}{2}-\Omega-\frac{i\Gamma_{g}}{2}+\frac{iV^2}{4}}(t_3+t_5e^{\left(iE+\Gamma_{g}\right)\frac{\xi_1}{2}})
\end{align}

As described in the main text, we calculate $e_B(x)$ in this case to verify whether the boundary condition holds in the solution given by bound state. It is concluded that, under the Markov approximation, the boundary condition that $e_B(x)$ satisfies is the same as the $e_F(x)$ solved by Bethe ansatz method.
\section{Derivation of the Eq.~(\ref{eq:Bk})}\label{App:B}
Since the propagation of two photons in the $o$-mode space is free, we only need to calculate the output states in the $e$-mode space. In this appendix, we provide a detailed derivation of Eq.~(\ref{eq:Bk}), with a particular focus on the bound state term $ B $. The S-matrix in the e-mode space $S_{ee}$ for two photons has been constructed in Eq.~(\ref{eq:eq54})
\begin{equation}
S_{ee} = \sum_{k \leq p}t_kt_p|F_{i,E}\rangle\langle F_{i,E}| + \sum_{E} t_B |B_{i,E}\rangle\langle B_{i,E}| .
\end{equation}
As described in main text, we consider the in-state $| S_{k_1, p_1} \rangle$ in the $e$-mode space first.
To analyze the momentum distribution of the output state in the $e$-mode space, we compute:
\begin{align}
\langle S_{k_2, p_2} |S_{ee} | S_{k_1, p_1} \rangle &= \sum_{k \leq p} t_k t_p \langle S_{k_2, p_2} | F_{i,E} \rangle \langle F_{i,E} | S_{k_1, p_1} \rangle \nonumber \\
&\quad + \sum_{E} t_B \langle S_{k_2, p_2} | B_{i,E} \rangle \langle B_{i,E} | S_{k_1, p_1} \rangle. \label{eq:Sk2}
\end{align}
The uncorrelated term in Equation~(\ref{eq:Bk}) comes from the first and second terms of the expression above, while the two-photon correlation term $B$ is contributed solely by the second term.
We will now evaluate each term on the right-hand side of this equation separately.

\subsection{First term of Eq.~(\ref{eq:Sk2})}
In this sub-appendix, we calculate the first term on the right side of Eq.~(\ref{eq:Sk2}).
Using the method described in the reference~\cite{Shen2007PRA} for natural atom, the result can be derived by  replacing $\Gamma$ with $\Gamma_g$ for the giant atom:

\begin{align}
&\sum_{k \leq p}t_k t_p \langle S_{k_2, p_2} | F_{i,E} \rangle \langle F_{i,E} | S_{k_1, p_1} \rangle \nonumber \\
&= \frac{1}{2} \sum_{k, p} t_k t_p \langle S_{k_2, p_2} | F_{i,E} \rangle \langle F_{i,E} | S_{k_1, p_1} \rangle \nonumber \\
&= \frac{1}{2} \int_{-\infty}^{+\infty} dE \int_{-\infty}^{+\infty} d\Delta \, \delta(E - E_1) \delta(E - E_2) t_k t_p  \nonumber \\
&\times\Bigg\{\Delta_1^{\prime}\times \Delta_2^{\prime}  - \left(\frac{\Gamma_g}{\pi}\right) \frac{4\Delta_1^2}{ 4\Delta_1^2 + \Gamma_g^2} \mathcal{P} \frac{\Delta_1^{\prime}}{\Delta_1^2 - \Delta_2^2}  \nonumber \\
& - \left(\frac{\Gamma_g}{\pi}\right) \frac{4\Delta_2^2}{ 4\Delta_2^2 + \Gamma_g^2} \mathcal{P} \frac{\Delta_2^{\prime}}{\Delta_2^2 - \Delta_1^2}  \nonumber \\
&+\left(\frac{\Gamma_g}{\pi}\right)^2 \frac{4\Delta^2}{4\Delta^2 + \Gamma_g^2} \mathcal{P} \frac{1}{\Delta_1^2 - \Delta_2^2} \mathcal{P} \frac{\Delta_1^2-\Delta_2^2}{(\Delta^2 - \Delta_1^2)(\Delta^2-\Delta_2^2)}  \Bigg\}, \label{eq:first}
\end{align}
where we have set $\Delta_{1(2)}^{\prime}=\delta\left(\Delta-\Delta_{1(2)}\right)+\delta\left(\Delta+\Delta_{1(2)}\right)$. Here, $ \mathcal{P} $ represents the Cauchy principal value integral. Since it involves integration with respect to $ \Delta $ and $ E $,
the product of the single photon transmission coefficients $ t_k t_p $ can be expressed in terms of $\Delta$ and $E$ as:
\begin{widetext}		
\begin{align}
t_k t_p &=
 \frac{2\Delta + (E - 2 \Omega) - i \Gamma_e^*}{2\Delta + (E - 2 \Omega) + i \Gamma_e}\times
\frac{2\Delta - (E - 2 \Omega) + i \Gamma_e^*}{2\Delta - (E - 2 \Omega) - i \Gamma_e},
\end{align}
where $2\Delta=k-p$ is the energy difference of two photons and $\Gamma_e$ is given in Eq.~(\ref{eq:Gamma_e}).
The first term in Eq.~(\ref{eq:first}) can be simplified to:
\begin{align}
t_{k_1} t_{p_1} & \delta(E_1 - E_2) [\delta(\Delta_1 - \Delta_2) + \delta(\Delta_1 + \Delta_2)] \nonumber \\
&= t_{k_1} t_{p_1} \delta(k_1 - k_2) \delta(p_1 - p_2)+ t_{k_1} t_{p_1} \delta(k_1 - p_2) \delta(p_1 - k_2).
\end{align}

The second and third terms in Eq.~(\ref{eq:first}) can be combined as:

\begin{align}
& \frac{\Gamma_g}{\pi} \Bigg( t_{k_2} t_{p_2} \frac{4 \Delta_2^2}{4 \Delta_2^2 + \Gamma_g^2} \mathcal{P} \frac{1}{\Delta_1^2 - \Delta_2^2}+ t_{k_1} t_{p_1} \frac{4 \Delta_1^2}{4 \Delta_1^2 + \Gamma_g^2} \mathcal{P} \frac{1}{\Delta_2^2 - \Delta_1^2} \Bigg) \delta(E_1 - E_2). \label{eq:sec}
\end{align}

For the fourth term in Eq.~(\ref{eq:first}), the integration can be calculated analytically by using a contour integral.
		
\begin{align}
& \frac{4i}{\pi} \Gamma_g^2 \bigg[ \frac{(\Gamma_e^* + \Gamma_e)(2E - i \Gamma_e^* + i \Gamma_e - 4 \Omega)(-iE + \Gamma_e + 2i \Omega)}{(4 \Delta_1^2 - (E + i \Gamma_e - 2 \Omega)^2)(-4 \Delta_2^2 + (E + i \Gamma_e - 2 \Omega)^2)(E^2 + \Gamma_g^2 + 2iE(\Gamma_e + 2i \Omega) - (\Gamma_e + 2i \Omega)^2)} \nonumber \\
&\quad + \frac{\Gamma_g (E - i \Gamma_e^* + i \Gamma_g - 2 \Omega)(iE + \Gamma_e^* + \Gamma_g - 2i \Omega)}{(4 \Delta_1^2 + \Gamma_g^2)(4 \Delta_2^2 + \Gamma_g^2)(iE + \Gamma_g - \Gamma_e - 2i \Omega)(-iE + \Gamma_g + \Gamma_e + 2i \Omega)} \bigg] \delta(E_1 - E_2). \label{eq:for}
\end{align}
		
\subsection{Second term of Eq.~(\ref{eq:Sk2})}
In this sub-appendix, we calculate the second term on the right side of Eq.~(\ref{eq:Sk2}). The overlap between $ |B_{i,E}\rangle $ and $ |S_{k_1, p_1}\rangle $ for the giant atom  can be obtained as~\cite{Shen2007PRA}:
\begin{align}
\langle B_{i,E} | S_{k_1, p_1} \rangle &= \frac{\sqrt{\Gamma_g}}{\sqrt{4 \pi}} \int_{-\infty}^{\infty} dx_c \int_{-\infty}^{\infty} dx \, e^{-iEx_c - \Gamma_g/2 |x|} \left[ \frac{1}{2 \pi} \frac{1}{\sqrt{2}} e^{iE_1 x_c} (e^{i\Delta_1 x} + e^{-i\Delta_1 x}) \right] \nonumber\\
&= \frac{\sqrt{\Gamma_g}}{\sqrt{4 \pi}} \frac{1}{2 \pi} \frac{1}{\sqrt{2}} \cdot 2 \pi \delta(E_1 - E) \int_{0}^{\infty} dx \, \left(e^{(i\Delta_1 - \Gamma_g/2)x} + e^{(-i\Delta_1 - \Gamma_g/2)x}\right) \nonumber\\
&= \frac{\sqrt{\Gamma_g}}{\sqrt{2 \pi}} \delta(E_1 - E) \left( \frac{-1}{i \Delta_1 - \Gamma_g/2} + \frac{-1}{-i \Delta_1 - \Gamma_g/2} \right) \nonumber\\
&= \frac{\sqrt{\Gamma_g}}{\sqrt{2 \pi}} \frac{4 \Gamma_g}{4 \Delta_1^2 + \Gamma_g^2} \delta(E_1 - E).
\end{align}
Thus, the bound state contribution to the second term of Eq.~(\ref{eq:Sk2}) is:
\begin{align}
\langle S_{k_2, p_2} | \left( \sum_{E} t_B |B_{i,E}\rangle \langle B_{i,E}| \right) | S_{k_1, p_1} \rangle &= \frac{\Gamma_g}{2 \pi} \int_{-\infty}^{\infty} t_B \frac{4 \Gamma_g}{4 \Delta_1^2 + \Gamma_g^2} \frac{4 \Gamma_g}{4 \Delta_2^2 + \Gamma_g^2} \delta(E - E_1) \delta(E - E_2) \nonumber\\
&= \frac{8 \Gamma_g^3}{\pi} t_{B1} \frac{1}{4 \Delta_1^2 + \Gamma_g^2} \frac{1}{4 \Delta_2^2 + \Gamma_g^2} \delta(E_1 - E_2), \label{eq:five}
\end{align}
where $t_B=t_6$ given in Eq.~(\ref{eq:t6}) and $t_{B1}$ can be obtained by replacing $E$ with $E_1$ in $t_6$.
In the Markovian regime where $ \Gamma_g \xi_1 = 0 $, the transmission coefficient $ t_B $ simplifies to:
\begin{align}
t_B &= \frac{E - 2 \Omega - i \Gamma_g - i \Gamma_e^*}{E - 2 \Omega + i \Gamma_g + i \Gamma_e}.
\end{align}
Thus, we have:
\begin{align}
\langle S_{k_2, p_2} | S_{ee} | S_{k_1, p_1} \rangle &= t_{k_1} t_{p_1} \delta(k_1 - k_2) \delta(p_1 - p_2) \nonumber\\
&\quad + t_{k_1} t_{p_1} \delta(k_1 - p_2) \delta(p_1 - k_2) + B \delta(E_1 - E_2),
\end{align}
where $ B $ is given by $ \text{Eq.~(\ref{eq:sec})} + \text{Eq.~(\ref{eq:for})} + \text{Eq.~(\ref{eq:five})} $. Under the Markov approximation, this simplifies to:
\begin{equation}
B = \frac{16 i \Gamma_g^2 (E_1 - 2 \Omega + i \Gamma_e)}{\pi \left[ 4 \Delta_1^2 - (E_1 - 2 \Omega + i \Gamma_e)^2 \right] \left[ 4 \Delta_2^2 - (E_1 - 2 \Omega + i \Gamma_e)^2 \right]}.
\end{equation}
which represents the correlated effect of the two photons.
\end{widetext}
\section{Derivation of S-matrix for giant atom with N coupling points}\label{App:N}
Similar to the case of giant atom with two coupling points, we set the eigenstate of $H_e$ for giant atom with $N$ coupling points $F^N(x_1,x_2)$ with the formalism of the Bethe ansatz as
\begin{align}\label{eq:eqD1}
	F^N_{l, E}(x_1, x_2) = A_l e^{ikx_1 + ipx_2} + B_l e^{ikx_2 + ipx_1},
\end{align}
with $l=1,\,\cdots\,,2N+1$ for a given energy $E$ of two photons. Since we only need to calculate the transmission coefficient $t^N_kt^N_p$, $2N+1$ is the minimum number in this process. For the giant atom with two coupling points, this process corresponds to $\mathcal{F}_1\rightarrow\mathcal{F}_2\rightarrow\mathcal{F}_3\rightarrow\mathcal{F}_5\rightarrow\mathcal{F}_6$ illustrated in Fig.~\ref{fig:fig3}.
Here, $F^N(x_1,x_2)$ is defined by plus the superscript $N$ in Eq.~(\ref{F}). And the subscript $l$ denotes the two-photon wave function $F^N(x_1, x_2)$ when two photons are in different regions, i.e.,  $F^N_{l}(x_1, x_2)$ denotes the wave function corresponding to the $l$th location of two photons.   For example, $F^N(x_1, x_2) \equiv{F}^N_{1,E}(x_1, x_2) $ when both photons locate at the left of the origin, i.e., $x_1<x_2<0$;   $F(x_1, x_2) \equiv{F}^N_{2N+1,E}(x_1, x_2) $ when two photons are in the right region of the position $\xi_{N-1}$,  i.e., $x_2>x_1>\xi_{N-1}$. The correspondence between two-photon wave function $F^N_{l,E}(x_1,x_2)$ and photon locations is summarized as follows:

For two photons located in the regions $x_1<0$, $ \xi_{l-2}<x_2<\xi_{l-1}$
\begin{align}\label{eq:D2}
	F^N(x_1,x_2)&=F^N_{l,E}(x_1,x_2),
\end{align}
where $1\leq l\leq N+1$, and we define $\xi_{-1}=-\infty$ and $\xi_{N}=+\infty$.

In the regions $\xi_{l-1}<x_1<\xi_l$, $\xi_{N-1}<x_2$
\begin{align}\label{eq:D3}
	F^N(x_1,x_2)=F^N_{l+N+1,E}(x_1,x_2)
\end{align}
where $1\leq l\leq N$ and  $\xi_N=+\infty$. Similar to the giant atom with two coupling points, we obtain the ratio $A_l/A_1$ and $B_l/B_1$ with $l=1,\,\cdots\,,2N+1$. In particular, we have the relation $A_{2N+1}/A_1=B_{2N+1}/B_1=t^N_kt^N_p$, where $t^N_k$ is given in Eq.~(\ref{eq:tkN}) and $t^N_p$ can be obtained by replacing $k$ with $p$. To obtain the ratio $A_1/B_1$, we formally provide the expression for $e_F^N(x)$.
For $x < 0$, we have
\begin{align}
	e_F^N(x) &= \frac{V \left(\sum_{j=1}^{N} A_{j} e^{i p \xi_{j-1}}\right)}{\sqrt{2} \left(p - \Omega + \frac{i V^2}{2 N}\right)} e^{i k x} \nonumber \\
	&\quad + \frac{V \left(\sum_{j=1}^{N} B_{j} e^{i k \xi_{j-1}}\right)}{\sqrt{2} \left(k - \Omega + \frac{i V^2}{2 N}\right)} e^{i p x}. \label{$}
\end{align}
For $x>\xi_{N-1}$, we have
\begin{align}
	e_F^N(x) &=\frac{V \left(\sum_{j=1}^{N}A_{j+N}e^{ip\xi_{j-1}}\right)}{\sqrt{2}\left(p-\Omega+\frac{iV^2}{2N}\right)}e^{ikx}\nonumber\\
	&\quad +\frac{V
		\left(\sum_{j=1}^{N}B_{j+N}e^{ip\xi_{j-1}}\right)}{\sqrt{2}\left(k-\Omega+\frac{iV^2}{2N}\right)}e^{ipx}
\end{align}
Using the boundary condition stated in Eq.~(\ref{eq:eq81}), we have the relation
\begin{align}
\frac{A_1}{B_1} = \frac{k - p - i \Gamma_N}{k - p + i \Gamma_N}.
\end{align}
where $\Gamma_N$ is given in Eq.~(\ref{eq:gammaN}). For the bound state, we can similarly get the coefficients $t_{l}$ for $l=1,\,\cdots\,,2N+1$, where we have set $t_1=1$ without loss of generality. Specially, we obtain the transmission coefficient $t_{2N+1}=t^N_B$ as given in Eq.~(\ref{eq:tEN}).

Moreover, for $e_B^N(x)$, we obtain formally for $x < 0$
\begin{equation}
	e_B^N(x) =  \frac{\sqrt{2}Ve^{(i E + \Gamma_N) x / 2}}{E-2\Omega + i \Gamma_{N} + \frac{i V^2}{ N}} \left( \sum_{j=1}^{N} t_j e^{(i E - \Gamma_N) \xi_{j-1} / 2} \right).
\end{equation}
For $x>\xi_{N-1}$,
\begin{equation}
	e_B^N(x) = \frac{\sqrt{2}Ve^{(i E - \Gamma_N) x / 2}}{E - 2\Omega - i \Gamma_{N} + \frac{i V^2}{ N}} \left( \sum_{j=1}^{N} t_{j+N} e^{(i E + \Gamma_N) \frac{\xi_{j-1}}{2}} \right),
\end{equation}
where $e^N_B(x)$ is the probability amplitude that there is one photon in the waveguide and the giant atom is in the excited state for the bound state case. Here, the superscript $N$ denotes the giant atom with $N$ coupling points similar to $e_F^N(x)$ stated in the main text. And $\Gamma_N$ is defined in Eq.~(\ref{eq:gammaN}). We substitute $e_B^N(x)$ into the boundary condition in Eq.~(\ref{eq:eq81}), and find the equation still holds when the Markov approximation is adopted similar to the giant atom with two coupling points.

\end{document}